\documentstyle[aps,multicol,epsf]{revtex}
\renewcommand{\theequation}{\arabic{section}.\arabic{equation}}

\begin{document}
\title{Vortex Pinning and the Non-Hermitian Mott Transition} 
\author{Raphael A. Lehrer and David R. Nelson} 
\address{Lyman Laboratory of Physics, Harvard University, Cambridge,
Massachusetts 02138}
\date{\today}
\maketitle

\begin{abstract}
The boson Hubbard model has been extensively studied as
a model of the zero temperature superfluid/insulator transition in $^4
\mbox{He}$ on periodic substrates.  It can also serve as a model for
vortex lines in superconductors with a magnetic field parallel to a
periodic array of columnar pins, due to a formal analogy between the
vortex lines and the statistical mechanics of quantum bosons.  When
the magnetic field has a component perpendicular to the pins, this
analogy yields a \textit{non-Hermitian} boson Hubbard model.  At
integer filling, we find that for small transverse fields, the
insulating phase is preserved, and the transverse field is
exponentially screened away from the boundaries of the superconductor.
At larger transverse fields, a ``superfluid'' phase of tilted,
entangled vortices appears.  The universality class of the transition
is found to be that of vortex lines entering the Meissner phase at
$H_{c1}$, with the additional feature that the direction of the tilted
vortices at the transition bears a nontrivial relationship to the
direction of the applied magnetic field.  The properties of the Mott
insulator and flux liquid phases with tilt are also discussed.
\end{abstract}

\begin{multicols}{2}

\section{Introduction}
\label{sec-intro}

The importance of interactions for vortex lines in superconductors has
been well known since Abrikosov's theory of the mixed phase of type II
superconductors.  The vortices form a regular triangular lattice due
to their repulsive interactions.  With the discovery of the cuprate
high-temperature superconductors, however, it became clear that
thermal fluctuations can play an important role in the behavior of the
flux lines arrays.~\cite{Blatter} The competition between
interactions, pinning, and thermal fluctuations gives rise to a wide
range of novel phenomena.  For example, it is now clear that the
triangular flux lattice melts via a first order
transition~\cite{Brezin+N} in clean materials, giving rise to an
entangled flux liquid~\cite{N88} over a significant part of the phase
diagram.

Here we investigate the behavior of an interacting system of flux
lines in a periodic array of columnar defects.  Traditionally,
columnar defects are made in high-temperature superconductors by
heavy-ion
bombardment.~\cite{Budhiani,Konczykowski,Civale,Gerhauser,Hardy} The
ions pass through the sample, leaving amorphous damage tracks in their
wake, which act very effectively as pinning sites for the flux lines,
provided the lines are aligned with the defects. Pinning is efficient
in this case because the dimensions of the flux lines are comparable
to those of the pins: both the vortices and the columnar defects are
extended over the length of the sample in one direction, and the
vortex core size ($\sim 20 \mbox{ } \AA$) is comparable to the width
of the damage tracks ($\sim 60 \mbox{ } \AA$).  By covering the sample
with a mask with a periodic array of holes through which the ions can
pass, a periodic array of pins (or clusters of pins) could be formed.

There are, however, other promising approaches to forming a periodic
array of columnar defects in which the artificial pinning centers are
able to accommodate multiple vortex lines per pin.  Such structures
have in fact already been made both in thick~\cite{Cooley} and thin
samples,~\cite{Baert} and one goal of this paper is to understand
effects of thermal fluctuations and entanglement when the materials
are very thick.  The work of Cooley \textit{et al}.~\cite{Cooley} is
of particular interest to us, because their samples can potentially be
used in two different ways to test our theories.  Their samples have a
regular lattice of pins within an ``island.''  The islands are
themselves spaced periodically through the sample.  We may use our
theory to predict behavior of the pins within a single island; or, we
may view each island as a large pinning center capable of
accommodating many vortices and apply our theories to that as well.
The lattices are triangular in either case.  We here describe effects
expected to be observed in square lattices of pinning sites; it is
straightforward but tedious to generalize our results to triangular
lattices as well.

Another motivation for examining flux lines in the presence of an
ordered array of columnar pins is to better understand computer
simulations such as those by Li and Teitel~\cite{Li+Teitel} and
others~\cite{simulations} which attempt to model flux lines in cuprate
superconductors.  These authors use an underlying discretization (such
as a stack of triangular lattices) for the simulations.  Although
layering along the \textit{c}-axis is an important feature of real
cuprates, the relatively coarse mesh of allowed sites for vortices in
each layer is clearly an artifact of the simulations.  As we shall
see, the presence of such periodicity can cause significant changes
(such as a transverse Meissner effect) in the behavior of the flux
lines.  The very existence of a periodic ``Mott insulator'' phase in
these simulations (see below) with its massive phonon modes and
infinite tilt modulus, is an artifact of the lattice discretization.
In the absence of a lattice, the Mott insulator phase would be
replaced by an Abrikosov crystalline phase with finite elastic
constants.  Our investigation can begin to delimit the circumstances
under which lattice approximations fail as models of real materials,
and to predict what kind of behavior might be observed when they do
fail.

As a third motivation, we hope that studies such as this will lead to
a better understanding of interacting flux lines with tilted field in
a \textit{random} array of columnar defects.~\cite{N+Vinokur,Hatano+N}
Although the extension of the standard theory of disordered
superfluids to vortex lines with tilted magnetic fields is
straightforward in principle~\cite{Hatano+N,Tauber+N} there are still
poorly understood aspects of this problem.

To understand how our model differs from conventional lattice boson
Hubbard models,~\cite{Fisher} consider the transfer matrix generated
by the classical partition function for wiggling vortex lines in $d +
1$ dimensions.  This transfer matrix is the exponential of a
$d$-dimensional non-Hermitian quantum Hubbard
Hamiltonian~\cite{N+Vinokur,Hatano+N}
\begin{eqnarray}
\hat{\cal H} & = & -t \sum_{\langle ij \rangle} \left( e^{{\bf h}
\cdot \hat{\bf e}_{ij}} \hat{a}_j^{\dag} \hat{a}_i + e^{ - {\bf h}
\cdot \hat{\bf e}_{ij}} \hat{a}_i^{\dag} \hat{a}_j \right) - \mu
\sum_j \hat{a}_j^{\dag} \hat{a}_j \nonumber \\
& & \mbox{} + \frac{U}{2} \sum_j \hat{a}_j^{\dag} \hat{a}_j \left(
\hat{a}_j^{\dag} \hat{a}_j - 1 \right).
\label{eq:hubbard}
\end{eqnarray}
The connection between this model and the physics of flux lines will
be reviewed in more detail in Sec.~\ref{sec-model}.  Here, we
simply note that the $\{ \hat{a}_i^{\dag}, \hat{a}_j \}$ are boson
(i.e., vortex) creation and destruction operators, the $\hat{\bf
e}_{ij}$ are unit vectors connecting nearest neighbor lattice sites on
a square lattice representing the positions of columnar pins in $d$
dimensions, and the first sum is over neighboring pairs of pins.  The
parameter $t$ controls vortex hopping between columns, $\mu = \mu
(H_z)$ is a chemical potential that depends on the component of the
external field $H_z$ parallel to the columns, and $U$ represents the
penalty for multiple vortices on a single defect.  The external
tipping field ${\bf H}_{\perp}$ determines the non-Hermitian hopping
asymmetry ${\bf h}$ via ${\bf h} = \phi_0 {\bf H}_{\perp} a_0/(4 \pi
k_B T)$, where $a_0$ is the lattice constant.  In the presence of this
non-Hermitian hopping, which would be impossible in conventional many
body quantum mechanics, the excitation spectrum can become complex.
The complex spectra which arise when $U = 0$ and $\mu$ varies randomly
from site to site has aroused considerable interest
recently.~\cite{Hatano+N,interest} In this paper, we study how this
non-Hermiticity affects excitations and other features of the Mott
insulator and superfluid phases.

Before discussing the interacting case, consider a system of {\em
noninteracting} flux lines in the presence of a periodic array of
columnar pins.  At very low temperatures in thin samples, each vortex
will be tightly bound to a particular pin, and fluctuations away from
the pinning sites will be small.  However, a flux line can gain
entropy by hopping from one defect to another.  For an infinitely
thick sample, the vortices will in fact be delocalized at {\em any}
nonzero temperature.  In the quantum analogy, this delocalization is a
consequence of the periodicity of the pinning potential and Bloch's
theorem.~\cite{Bloch} Although the distance in the $\hat{\bf
z}$-direction between intercolumn hops diverges as the temperature is
lowered to zero,~\cite{Balents+N} we will assume sufficiently thick
samples and high temperatures so that an individual vortex hops
repeatedly as it traverses the sample.  Such delocalized states are
easy to tilt and will have a finite linear resistance, because
transport in the presence of a Lorentz force will be aided by finite
concentration of kinks in equilibrium.~\cite{N+Vinokur} Hopping from
one columnar pin to another \textit{cannot} be neglected at finite
temperatures in thick samples, even if such events are relatively
rare.

We now discuss qualitatively the interacting case.  Consider an
external magnetic field $H_z$ such that there is exactly one flux line
per columnar pin, i.e., $B_z = \phi_0 / a_0^2$.  If the pins are
strong enough, the ground state will be a square lattice of vortices
rather than the usual triangular array that is observed in the absence
of pins.  Thermal fluctuations allow the lines to make short
excursions away from the columnar pins which lowers the effective
strength of the pins.~\cite{N+Vinokur} More substantive changes arise
if thermal disorder also leads to hops from one pin to another, as for
the noninteracting flux lines discussed above.  However, the
interactions inhibit hopping, due to the extra energy cost of double
occupancy.~\cite{Note} Although hopping increases the entropy per unit
length, this is insufficient to overcome the energy barrier at low
temperatures.  Above a certain critical temperature $T_c$, however,
the entropy gain dominates, and the flux lines can hop freely from pin
to pin.  In the quantum analogy, this phase change is the boson Mott
transition~\cite{Fisher} between a $2d$ Mott insulator (square vortex
lattice) and a $2d$ superfluid (entangled vortex liquid).  The
temperature $T$ for the vortices plays the role of $\hbar$ for the
bosons.  (See Sec.~\ref{sec-model}.)

\begin{minipage}[t]{3.2in}
\vspace{0.1in}
\epsfxsize=2.2in
\hfill \epsfbox{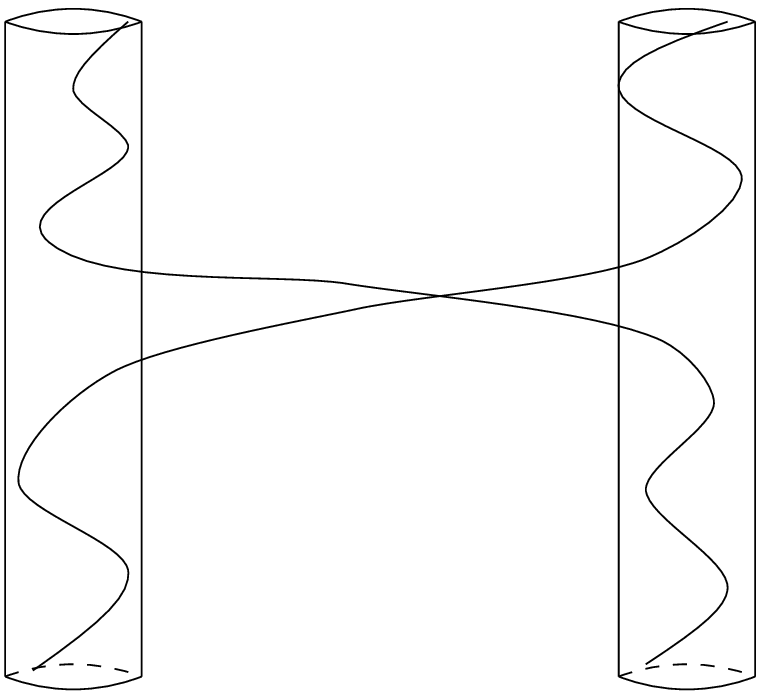} \hfill
\begin{small}

FIG.\ 1. 
An example of a correlated hopping event.  Such hops
delocalize flux lines, but cannot mediate vortex transport or tilt.
\end{small}
\vspace{0.2in}
\end{minipage}

\textit{Correlated} hopping events such as are shown in Fig.\ 1 can
lower the entropy without forcing two flux lines to occupy the same
column, and always delocalize lines in sufficiently thick samples even
in the crystalline phase.  However, such defects cannot mediate vortex
transport or tilt---if one flux line moves to the right, another must
move to the left, so there is no net change.  Thus it will not be
until the temperature rises above the critical temperature mentioned
above that there will be any linear resistance or response to a
perpendicular magnetic field.

Consider now raising the applied magnetic field ${\bf H} = H_z {\bf
\hat{z}}$, always keeping it parallel to the column direction.  In
the absence of columnar pins, the lattice spacing would simply shrink;
however, if the pins are strong enough, a change in the lattice
constant would sacrifice the energy benefit of commensurability with
the lattice of pinning sites.  An alternative response is the
introduction of extra vortices into the commensurate lattice.
Depending on the pinning strength, the additional vortices will either
doubly occupy certain columnar pins, or else occupy interstitial
sites; either possibility costs an extra energy cost per unit length.
We assume for simplicity that the cost of double occupancy is
lower~\cite{Note}, and call the resulting extra vortex a ``particle''
excitation.  This ``particle'' has an energy per unit length of order
the cost of double occupancy; however, it can gain entropy by hopping
from pin to pin, and thus lower its \textit{free energy}.  In fact,
its free energy can become negative.  Following the treatment of
interstitials and vacancies in a conventional Abrikosov lattice in
Ref.~\cite{Frey+N+F}, an estimate of the boundary above which
particles are favored arises from approximating the partition function
of a single extra vortex by $4^{L/l_z} \exp (- \epsilon_p L / T)$,
where $\epsilon_p = \epsilon_p (H_z)$ is the energy cost per unit
length of a particle, $L$ is the length in the $z$ direction, and
$l_z$ is the distance between hops of the particle from one site to
another.  If $E_p$ is the extra energy of an particle in zero external
magnetic field $H$, then $\epsilon_p (H) = E_p - \phi_0 H_z/4 \pi$.
The factor of 4 comes from the four directions available for hopping
on a square lattice.  Thus the free energy of the particle will be $F
\approx \epsilon_p L - T (L/l_z) \ln 4$, which becomes negative above
a temperature
\begin{equation}
T_d (H_z) = \frac{\epsilon_p (H_z) l_z}{\ln 4}.
\end{equation}
Above this temperature, particles are favorable; below it, however,
the system cannot respond to a change in $H_z$, and the extra magnetic
field is screened out.  Similar screening arises if one attempts to
produce vacancies or ``holes'' by reducing the magnetic field, with
$F_h \approx \epsilon_h L - T (L/l_z) \ln 4$, where $\epsilon_h$ is
the hole energy in the absence of an external magnetic field. Thus for
a range of applied fields $H_z$, the actual field $B_z$ (as measured
by the vortex density) will be locked in at the value $B_z = \phi_0 /
a^2$, much as in the Meissner effect where the field is locked in at
$B = 0$ for a range of applied $H_z$.  Similar arguments show that if
we apply a small field ${\bf H}_{\perp}$ transverse to the columns,
isolated flux lines will not be able to tilt in response to it, and
the field will again be screened out, i.e., ${\bf B}_{\perp} = {\bf
0}$.~\cite{N+Vinokur}

In the rest of the paper, we examine the phase transitions described
qualitatively above.  The phase diagram is shown in Fig.\ 2.  The
``commensurate'' Meissner-like phase discussed above for vortex matter
is called a ``Mott insulator'' in the context of quantum bosons.  With
${\bf H}_{\perp} = {\bf 0}$, this problem has been studied by Fisher
\textit{et al}.~\cite{Fisher} in the context of real quantum bosons on
a lattice.  In the presence of a transverse field, however, the formal
analogy between the statistical mechanics of vortex matter and
(2+1)-dimensional quantum mechanics leads to the
\textit{non-Hermitian} model of Eq.\
(\ref{eq:hubbard}).~\cite{Hatano+N}

\begin{minipage}[t]{3.2in}
\vspace{0.1in}
\epsfxsize=3.2in
\epsfbox{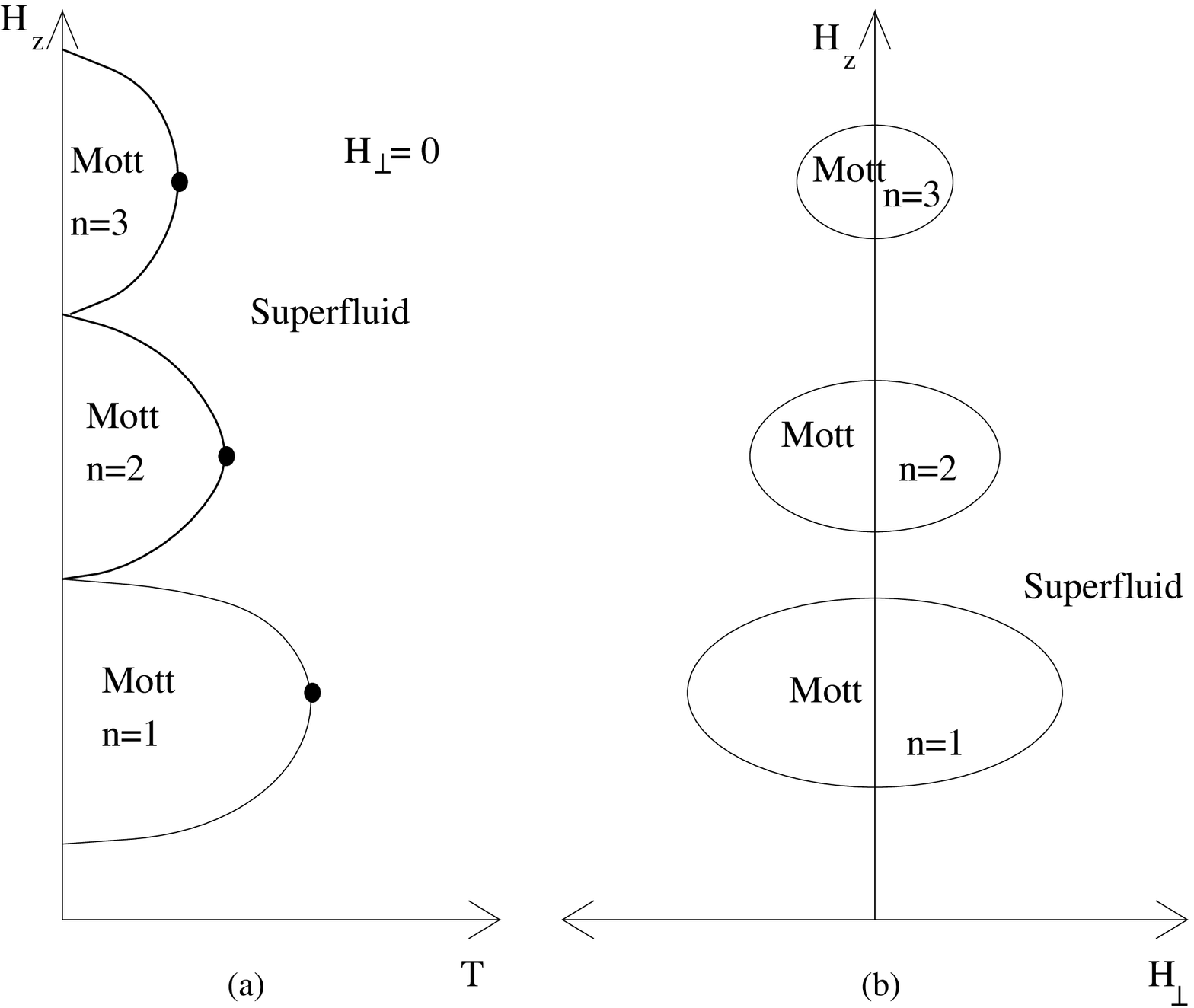}
\begin{small}
FIG.\ 2. 
(a) Schematic phase diagram for ${\bf H}_{\perp} = 0$.  The
circles at the tips of the Mott lobes are the multicritical points.
(b) Schematic phase diagram at fixed temperature $T$ below the
multicritical points in (a).
\end{small}
\vspace{0.2in}
\end{minipage}

In Sec.~\ref{sec-model}, we describe in more detail our model of
interacting vortex lines.  We study the Mott insulator to flux liquid
phase transition induced by a perpendicular field in
Sec.~\ref{sec-transition}, using mean field theory and a more
sophisticated renormalization group treatment.  We argue that in the
presence of a transverse field, the transition is characterized by the
penetration of ``tilted defects,'' which are particles or holes with
net average motion in response to the transverse field, as shown in
Figs.\ 3 and 4.  More support is given for this picture in
Sec.~\ref{sec-Mott}, where the low temperature Mott insulator phase is
examined.  We determine excitation gaps for hopping in
\begin{minipage}[t]{3.2in}
\vspace{0.1in}
\epsfxsize=3.2in
\epsfbox{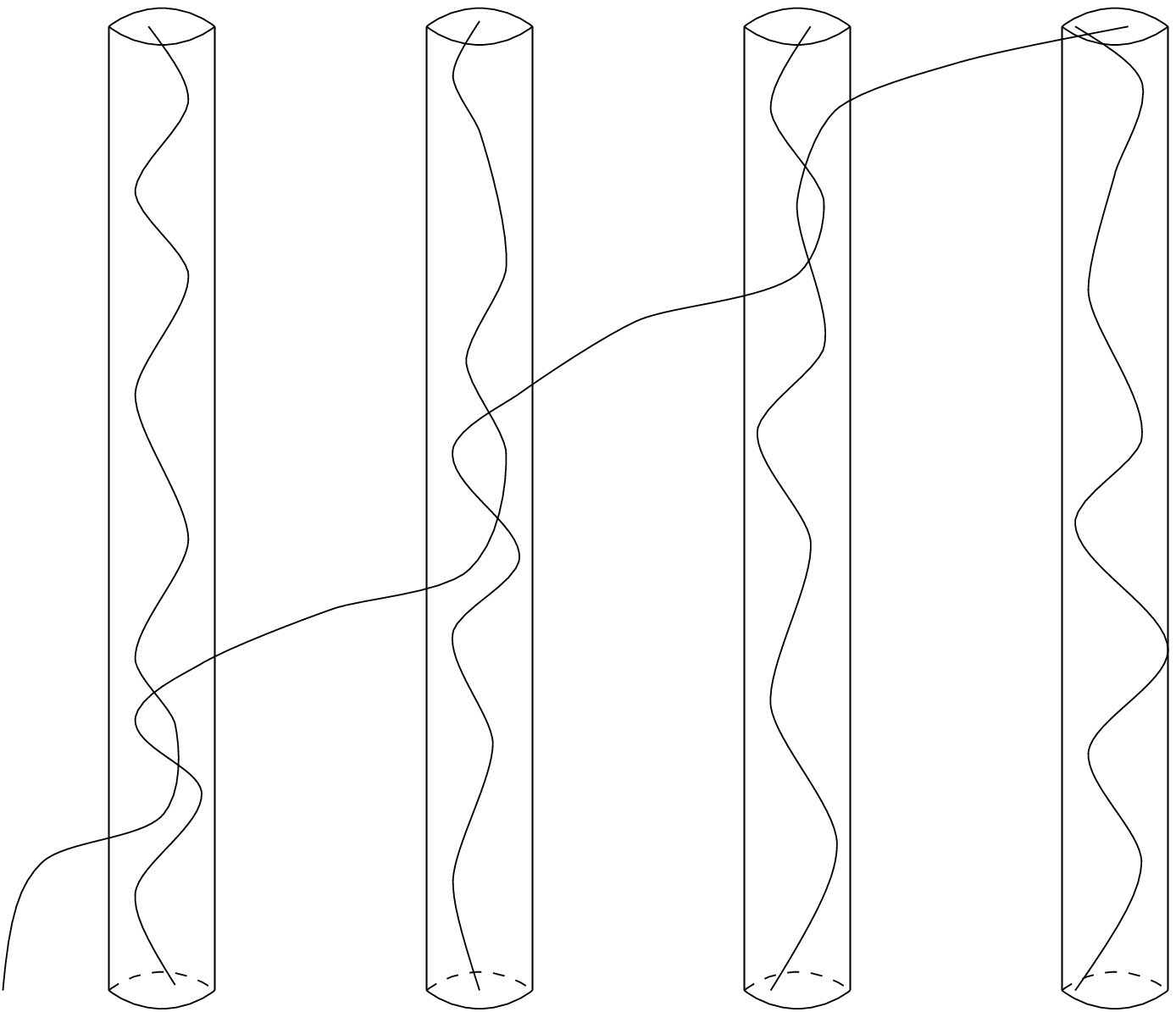}
\begin{small}
FIG.\ 3. 
An example of a right-moving tilted particle defect.
\end{small}
\vspace{0.2in}
\end{minipage}

\noindent the presence of the non-Hermitian perturbation.  We also
calculate a quantity analogous to the London penetration depth in the
Meissner phase, which is the length over which screening of the
transverse magnetic field becomes effective near the boundaries over
the sample.  In Sec.~\ref{sec-super}, we examine properties of the
superfluid phase, particularly the way we expect the magnetic field to
penetrate the sample near the transition, and the excitation spectrum.

\begin{minipage}[t]{3.2in}
\vspace{0.1in}
\epsfxsize=3.2in
\epsfbox{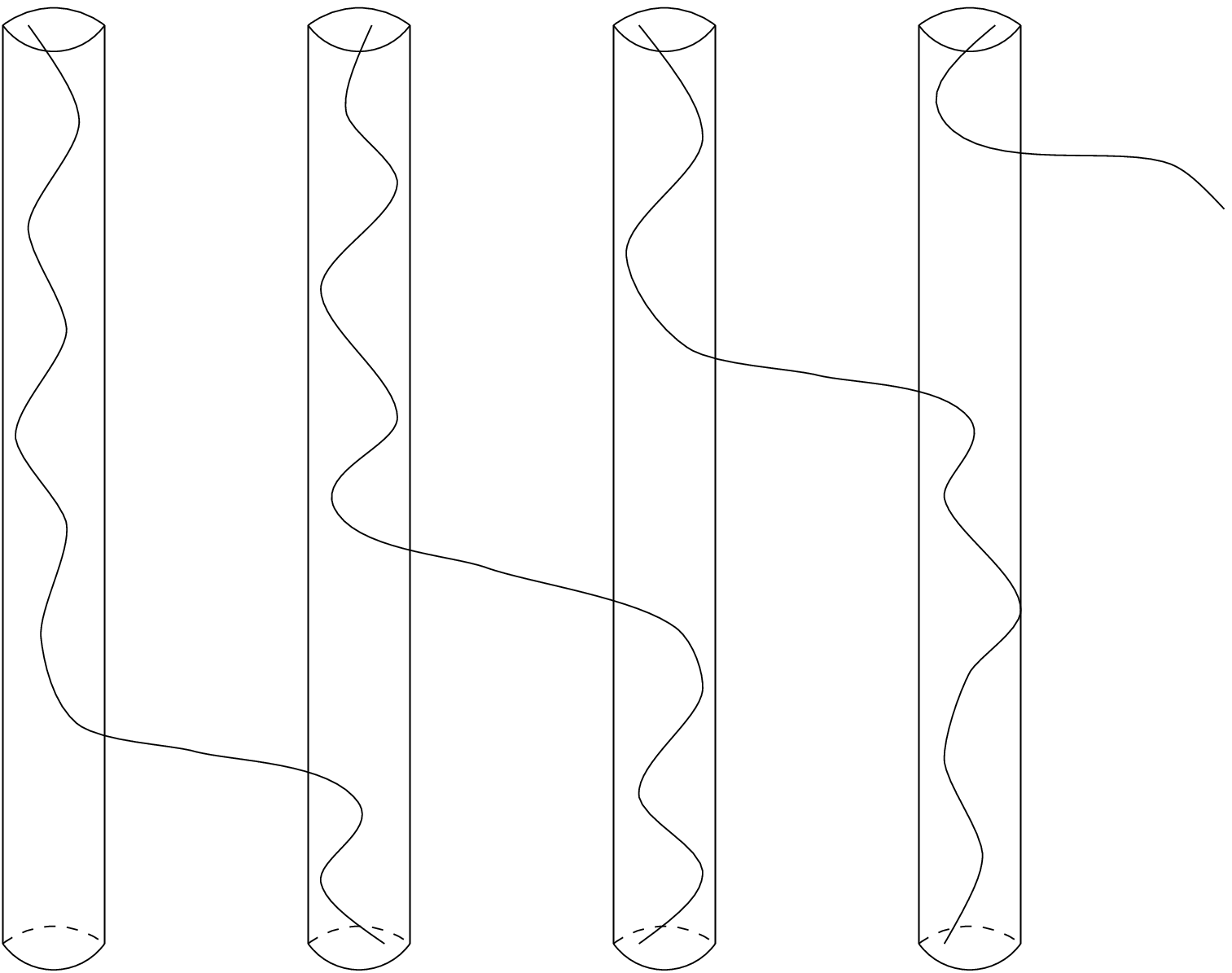}
\begin{small}
FIG.\ 4. 
An example of a right-moving tilted hole defect.
\end{small}
\vspace{0.2in}
\end{minipage}

\section{Model}
\label{sec-model}
\setcounter{equation}{0}

We begin with a model free energy for $N$ vortex lines in a sample of
thickness $L$ in the $\hat{\bf z}$ direction (perpendicular to the
$\mbox{CuO}_2$ planes) in the presence of a periodic square array of
columnar defects aligned in the $\hat{\bf z}$ direction with nearest
neighbor separation $a$.  We allow an external magnetic field ${\bf
H}_{\perp}$ transverse to the defects as well as the usual parallel
field ${\bf H}_{\parallel}$.  The energy then reads~\cite{N+Vinokur}
\begin{eqnarray}
\label{eq:model}
\displaystyle
F_N & = & \frac{1}{2} \tilde{\epsilon}_1 \sum_{\mu = 1}^{N} \int_0^L
\left| \frac{d {\bf r}_{\mu}(z)}{dz} \right|^2 dz \nonumber \\
\displaystyle
& & \mbox{} + \frac{1}{2} \sum_{\mu \neq \nu} \int_0^L V(| {\bf r}_{\mu}(z) -
{\bf r}_{\nu}(z)|) dz \nonumber \\
\displaystyle
& & \mbox{} + \sum_{\mu = 1}^{N} \int_0^L V_D[{\bf r}_{\mu}(z)] dz
\nonumber \\
\displaystyle
& & \mbox{} - \frac{{\bf H}_{\perp} \phi_0}{4\pi} \cdot \sum_{\mu = 1}^{N}
\int_0^L \frac{d {\bf r}_{\mu}(z)}{dz} dz,
\end{eqnarray}
with
\begin{equation}
V_D[{\bf r}_{\mu}(z)] = \sum_{i=1}^M V_1[|{\bf r}_{\mu}(z) - {\bf
R}_i|].
\end{equation}
Greek letters denote the flux lines while Roman letters denote the
columnar pins.  $V(r)$ is the (repulsive) interaction potential
between flux lines, which can be taken local in $\hat{\bf z}$, while
$V_1$ describes the (attractive) interaction between a flux line and a
columnar defect.  The parameter $\tilde{\epsilon}_1$ in the kinetic
energylike term arises from the tilt energy of the lines.  This term
is a quantitatively correct approximation to the tilt energy provided
\begin{equation}
\left| \frac{d {\bf r}_{\mu} (z)}{dz} \right|^2 \ll M_z/M_{\perp},
\end{equation}
where $M_z/M_{\perp} \gg 1$ is the mass anisotropy in the underlying
Ginzburg-Landau theory.  That same criterion insures that an
interaction local in $z$ is a good approximation.~\cite{N+Vinokur}

There is a useful formal correspondence between this flux line problem
and fictitious quantum-mechanical bosons in two
dimensions.~\cite{N88,N+Vinokur}.  The temperature $T$ plays the role
of Planck's constant $\hbar$, the bending energy $\tilde{\epsilon}_1$
plays the role of the boson mass $m$, and the length of the sample $L$
corresponds to $\beta \hbar$ for the bosons.  (See Table I for a
summary.)  However, while real bosons have \textit{periodic} boundary
conditions in imaginary time, \textit{free} boundary conditions are
more appropriate to the flux line system.  As we take the
thermodynamic limit $L \rightarrow \infty$ (corresponding to the zero
temperature limit for the fictitious bosons), the partition function
is dominated by the ground state, and the statistical mechanics is
insensitive to the boundary conditions.  Even for finite $L$, the only
states which contribute to the partition function of the vortex lines
are symmetrical boson states.~\cite{N88} We defer consideration of
finite size effects to Sec.~\ref{sec-boundary}, while in the rest of
this paper we use periodic boundary conditions for convenience and
take the limit $L \rightarrow \infty$.

\begin{minipage}[t]{3.2in}
\vspace{0.1in}
\begin{tabular}{|c|c|} 
\hline Vortex lines & Bosons \\ \hline \hline
$\tilde{\epsilon}_1$ &  $m$ \\ \hline
$k_B T$ & $\hbar$ \\ \hline
$L_z$ & $\beta \hbar$ \\ \hline
$H_z \phi_0 / 4 \pi - \tilde{\epsilon}_1$ & $\mu$ \\ \hline
${\bf H}_{\perp} \phi_0 a_0 / 4 \pi k_B T$ & ${\bf h}$ \\ \hline
Meissner-like commensurate phase & Mott insulator \\ \hline
Flux liquid  & Superfluid  \\ \hline 
\end{tabular}
\vspace{0.1in}
\begin{small}
TABLE I.
Detailed correspondence of the parameters of the flux line
system with the parameters of the two-dimensional boson system.
\end{small}
\vspace{0.2in}
\end{minipage}

As emphasized in Ref.~\cite{Hatano+N}, the transverse field ${\bf
H}_{\perp}$ plays the role of an \textit{imaginary} vector potential
in the boson system.  Thus, understanding the flux line system in the
presence of a transverse field is equivalent to understanding the
quantum mechanics of \textit{non-Hermitian} two-dimensional bosons
with interactions and a periodic potential.

In the limit of a very strong columnar pinning potential, we can
replace Eq.\ (\ref{eq:model}) by a simplified tight binding model for
the vortex lines~\cite{N+Vinokur,Hatano+N,Note} such that, written in
terms of the fictitious boson operators, the grand canonical partition
function is
\begin{equation}
{\cal Z} = \mbox{Tr} \left\{ e^{- \beta L \hat{\cal H}} \right\},
\label{eq:Z}
\end{equation}
where $\hat{\cal H}$ given by Eq.\ (\ref{eq:hubbard}) and $L$ is the
sample thickness.  The non-Hermitian terms in the effective
Hamiltonian reflect a preference for the bosons to hop in the
direction of the transverse field.  We have neglected interactions
between flux lines that are not on the same site.  This should give
reasonable results near filling fraction $1$; however, if we wanted to
describe filling fraction $1/2$, for example, where we might expect
the ground state to be a checkerboard arrangement of the flux lines,
we would need to include further neighbor interactions.  Further
neighbor interactions would also be required to treat the lattice
models of melting studied numerically in Refs.~\cite{Li+Teitel}
and~\cite{simulations}.

The parameters of this model, expressed in the original flux line
language are~\cite{N+Vinokur,Hatano+N}
\begin{eqnarray}
\label{eq:mu}
\mu & = & \frac{{\bf H}_z \phi_0}{4\pi} - \epsilon_1, \\
\label{eq:h}
{\bf h} & = & \frac{{\bf H}_{\perp} \phi_0 a_0}{4\pi k_B T}, \\
\label{eq:U}
U & \approx & 2 \epsilon_0 \ln (H_{c2}/H), \\
\label{eq:t}
t & \approx & 2 \left( \frac{2}{\pi} \right)^{1/2}
\frac{V_0}{\sqrt{E_{ij}/T}} e^{-E_{ij}/T},
\end{eqnarray}
with $V_0$ the strength of the pin and $E_{ij} = \sqrt{2
\tilde{\epsilon}_1 V_0} a_0$ the WKB exponent for tunneling from one pin
to its neighbor.  The Hubbard $U$ term in Eq.\ (\ref{eq:hubbard})
contributes an energy $\frac{1}{2} U n (n-1)$ when $n$ flux lines are
on a single site.  It represents the difference between the energy of
a single $n$-fold vortex quantum ($\sim n^2$) and the energy of $n$
well-separated unit vortices.  In this tight-binding limit, a Mott
insulator phase exists not only at filling fraction $1$, but at any
integer filling fraction.  We will consider Mott insulator phases with
general filling fraction $n$.  Integer filling fraction with $n \geq
2$ would arise as an approximation to samples irradiated through a
mask of holes large enough to allow multiple columnar defects at every
lattice site.

\section{Transition and Renormalization Group}
\label{sec-transition}
\setcounter{equation}{0}

Unless $\mu/U$ is exactly an integer [corresponding to the points
where two Mott lobes are tangent in Fig.\ 2(a)], it is
expected~\cite{Fisher} that the equivalent quantum system will be in a
Mott insulator phase with a well defined integer site occupation
number $n$ for sufficiently weak hopping parameter $t$.  As discussed
in Sec.~\ref{sec-intro}, the ``Mott insulator'' corresponds to a
Meissner-like vortex phase, which resists changes in the number of
vortices per site.  As is evident from Eq.\ (\ref{eq:t}), small $t$
corresponds to low temperatures in the vortex system.  Large hopping
leads eventually to a 2nd order phase transition to a superfluid
phase.  This ``superfluid'' is a lattice analogue of an entangled flux
liquid for vortex matter.  In this section, we first mention a simple
mean field theory due to Sheshadri, \textit{et al}.~\cite{Sheshadri}
The detailed application of this mean field theory to the
non-Hermitian boson Hubbard model is described in
Appendix~\ref{sec-mft}.  We then use a more systematic approach to
study this transition which will allow us to determine the
universality class of the transition.  We defer a detailed discussion
of the phases themselves until later sections.

\subsection{Mean field theory}

One simple way of accessing the mean field theory associated with Eq.\
(\ref{eq:hubbard}) is with an ansatz that decouples the hopping term
to give a single-site Hamiltonian via the replacement~\cite{Sheshadri}
\begin{equation}
\hat{a}_i^{\dag} \hat{a}_j \rightarrow \langle \hat{a}_i^{\dag} \rangle
\hat{a}_j + \hat{a}_i^{\dag} \langle \hat{a}_j \rangle - \langle
\hat{a}_i^{\dag} \rangle \langle \hat{a}_j \rangle.
\end{equation}
In the remainder of this section, however, we follow a different
approach that allows a more convenient treatment of fluctuation
effects.  The decoupling method, however, is conceptually and
computationally simpler, so we present it in Appendix~\ref{sec-mft}.

\subsection{Hubbard-Stratanovich transformation}

We here transform the tight binding partition function of Eq.\
(\ref{eq:Z}) and Eq.\ (\ref{eq:hubbard}) into a field theoretic form
more convenient for calculation following a method used by Fisher,
\textit{et al}.~\cite{Fisher}, and earlier by Doniach~\cite{Doniach},
to treat a quantum model of coupled arrays of Josephson junctions.  We
begin by separating the Hamiltonian $\hat{\cal H}_N$ of Eq.\
(\ref{eq:hubbard}) into the single site piece $\hat{\cal H}_0$ and the
off-site hopping term $\hat{\cal H}_1$: $\hat{\cal H}$ = $\hat{\cal
H}_0$ + $\hat{\cal H}_1$, with
\begin{eqnarray}
\displaystyle
\label{eq:h0}
\hat{\cal H}_0 & = & \sum_i \left[ \frac{U}{2} \hat{n}_i(\hat{n}_i -
1) - \mu \hat{n}_i \right], \\
\displaystyle
\hat{\cal H}_1 & = & - \frac{1}{2} \sum_{i,j} [J_{ij} \hat{a}_i^{\dag}
\hat{a}_j + J_{ji} \hat{a}_j^{\dag} \hat{a}_i],
\label{eq:h1}
\end{eqnarray}
where $\hat{n}_i = \hat{a}_i^{\dag} \hat{a}_i$ is the number operator
and $J_{ij}$ is the hopping matrix element between sites $i$ and $j$:
\begin{equation}
J_{ij} = \left\{ \begin{array}{ll}
te^{{\bf h \cdot \hat{e}}_{ij}} & \mbox{if $i,j$ nearest neighbors} \\
0 & \mbox{otherwise}.
\end{array} \right.
\label{eq:transverse}
\end{equation}

The generalization of the transformations of Refs.~\cite{Doniach}
and~\cite{Fisher} to the non-Hermitian case presents no difficulties.
Their effect is to allow the grand canonical partition function of
Eq.\ (\ref{eq:Z}) to be rewritten as a functional integral over a set of
complex functions $\{ \psi_i (\tau) \}$,
\begin{equation}
{\cal Z} = {\cal Z}_0 \int \prod_i {\cal D} \psi_i(\tau) {\cal D}
\psi_i^*(\tau) \exp [-S(\psi)],
\label{eq:Z2}
\end{equation}
with an effective action
\begin{eqnarray}
\displaystyle
S(\psi) & = & \beta \sum_{i,j} \int_0^{L} d \tau (J^{-1})_{ij}
\psi_i^*(\tau) \psi_j(\tau) \nonumber \\
\displaystyle
& & \mbox{} - \sum_i \ln \left\langle T_{\tau} \exp \left[ \beta
\int_0^{L} d \tau \{ \psi_i(\tau) \hat{a}_i^{\dag}(\tau)
\right. \right. \nonumber \\
\displaystyle
& & \hspace{1.2in} \left. \mbox{} + \psi_i^*(\tau) \hat{a}_i(\tau) \}
\Bigg] \right\rangle_0,
\label{eq:S}
\end{eqnarray}
where $T_{\tau}$ is the imaginary-time ordering operator,
\begin{eqnarray}
{\cal Z}_0 & = & \mbox{Tr} \left\{ e^{-\beta L \hat{\cal H}_0}
\right\}, \\
\langle \bullet \rangle_0 & = & \frac{1}{{\cal Z}_0} \mbox{Tr} \left\{
e^{-\beta L \hat{\cal H}_0} \bullet \right\},
\end{eqnarray}
and the imaginary time dependence of operators is given by, e.g.,
\begin{equation}
\hat{a}_i(\tau) = e^{\beta \tau \hat{\cal H}_0} \hat{a}_i e^{- \beta
\tau \hat{\cal H}_0}.
\end{equation}
We can see the equivalence of Eq.\ (\ref{eq:Z2}) with Eq.\
(\ref{eq:Z}) by integrating out the $\psi$ fields to obtain
\begin{equation}
{\cal Z} = {\cal Z}_0 \left\langle T_{\tau} \exp \left[ - \beta
\int_0^{L} d \tau \hat{\cal H}_1(\tau) \right] \right\rangle_0,
\end{equation}
which is just the familiar interaction representation of Eq.\
(\ref{eq:Z}).

\subsection{Universality classes}
\label{sec-universality}

Since $\langle \psi_i \rangle$ is an order parameter for
superfluidity~\cite{Doniach,Fisher} we can study the transition to a
state with $\langle \psi_i \rangle \neq 0$ by expanding Eq.\
(\ref{eq:S}) in powers of $\psi$.  For now, we keep only the most
relevant terms in a long-wavelength approximation, and obtain
\begin{eqnarray}
\displaystyle
S(\psi) \approx && \beta a_0^2 \int \frac{d^2 {\bf k} d
\omega}{(2\pi)^3} |\psi({\bf k}, \omega)|^2 [ r - i c_{\mu} \omega + i
{\bf c}_{\bf h} \cdot {\bf k} \nonumber \\
\displaystyle
& & \mbox{} + c' \omega^2 + D_{ij} k_i k_j ] + u \int d^2 {\bf r} d
\tau |\psi({\bf r}, \tau)|^4,
\label{eq:expandquad}
\end{eqnarray}
as shown in Appendix~\ref{sec-expansion}.  This formula is applicable
to a Mott insulator phase with $n$ particles per site (i.e. $n - 1 <
\mu /U < n$).  The values of the parameters $r(\mu, T)$, $c'$, $u$,
and the matrix $D_{ij}$ are given in Appendix~\ref{sec-expansion},
while
\begin{eqnarray}
\label{eq:cmu}
c_{\mu} & = & k_B T \left( \frac{n+1}{E_p^2} - \frac{n}{E_h^2}
\right), \\
{\bf c}_{\bf h} & = & \frac{a_0}{2t} \frac{\hat{\bf x} \sinh h_x +
\hat{\bf y} \sinh h_y}{[\cosh h_x + \cosh h_y]^2},
\label{eq:ch}
\end{eqnarray}
where
\begin{eqnarray}
\label{eq:barepart}
E_p & = & U n - \mu, \\
E_h & = & \mu - U (n-1)
\label{eq:barehole}
\end{eqnarray}
are (respectively) the energies of ``particles'' and ``holes''
superimposed on a Mott insulator state with a fixed value of $n$, in
the absence of hopping.

Note that $c_{\mu}$ and ${\bf c}_{\bf h}$ can be positive, negative,
or zero.  In the case examined in Fisher \textit{et al}., there is no
transverse field, i.e., ${\bf c}_{\bf h} = 0$.  If $c_{\mu} = 0$ as
well, then the universality class of the transition which occurs for
$r \approx 0$ is that of the \textit{XY} model in three dimensions,
with nontrivial critical exponents due to fluctuations.  When $c_{\mu}
> 0$, the transition at $r \approx 0$ is characterized instead by mean
field exponents with logarithmic
corrections.~\cite{Fisher,Fisher+Hohenberg} This is also the
universality class which describes the penetration of thermally
excited vortex lines from the Meissner phase near $H_{c1}$.~\cite{N88}
Near this transition, extra vortex lines enter into the sample and
behave (at long length scales) like a dilute gas of weakly interacting
bosons.  When $c_{\mu} < 0$, the universality class is again like flux
penetration near $H_{c1}$, with the magnetic field reversed.  Now,
holes enter the sample and behave (at long wavelengths) like a dilute
system of linelike excitations in the negative $\hat{\bf z}$
direction.

When a transverse field is applied, ${\bf c}_{\bf h} \neq 0$, and we
see from Eq.\ (\ref{eq:expandquad}) that the transition again
resembles the physics of dilute linelike excitations near $H_{c1}$,
but in a direction given by the three dimensional vector
\begin{equation}
\label{eq:angle}
{\bf v} = {\bf c}_{\bf h} + c_{\mu} \hat{\bf z}.
\end{equation}
When $r(\mu, T) \lesssim 0$, tilted holes or extra vortices enter the
sample, at an angle determined by Eq.\ (\ref{eq:angle}).  See Figs.\
3 and 4 for illustrations of tilted particle and hole defects, and see
Fig.\ 5 for the range of parameter space in which we expect each type
of defect.  The renormalization group calculation that describes the
transition, including effects of the underlying square columnar
lattice of preferred sites, is given in Appendix~\ref{sec-rg}.

We confirm this interpretation of the physics when we examine the Mott
insulator and superfluid phases in Secs.~\ref{sec-Mott}
and~\ref{sec-super}.  In the Mott insulator, we calculate the energies
per unit length of defects constrained to have a specific average
direction, and find that only those oriented in direction $\hat{\bf
v}$ have vanishing energies per unit length as we approach the
transition.  Thus, near the transition, only defects in this specific
orientation will have favorable energy and proliferate.  In the
superfluid phase, we calculate the magnetic field change $\delta {\bf
B}$ in excess of the field ${\bf B}_0$ locked into the Mott insulator
phase, and find that it lies in the $\hat{\bf v}$ direction as well,
corroborating our assertion that the defects penetrate at an average
angle
\begin{eqnarray}
\label{eq:theta}
\theta_v & = & \tan^{-1} \left( \frac{|{\bf c_h}|}{c_{\mu}} \right), \\
\phi_v & = & \tan^{-1} \left( \frac{c_{{\bf h}_y}}{c_{{\bf h}_x}} \right),
\label{eq:phi}
\end{eqnarray}
where $\theta_v$ and $\phi_v$ are polar angles with respect to the
$\hat{\bf z}$ axis.

\begin{minipage}[t]{3.2in}
\vspace{0.1in}
\epsfxsize=3.2in
\epsfbox{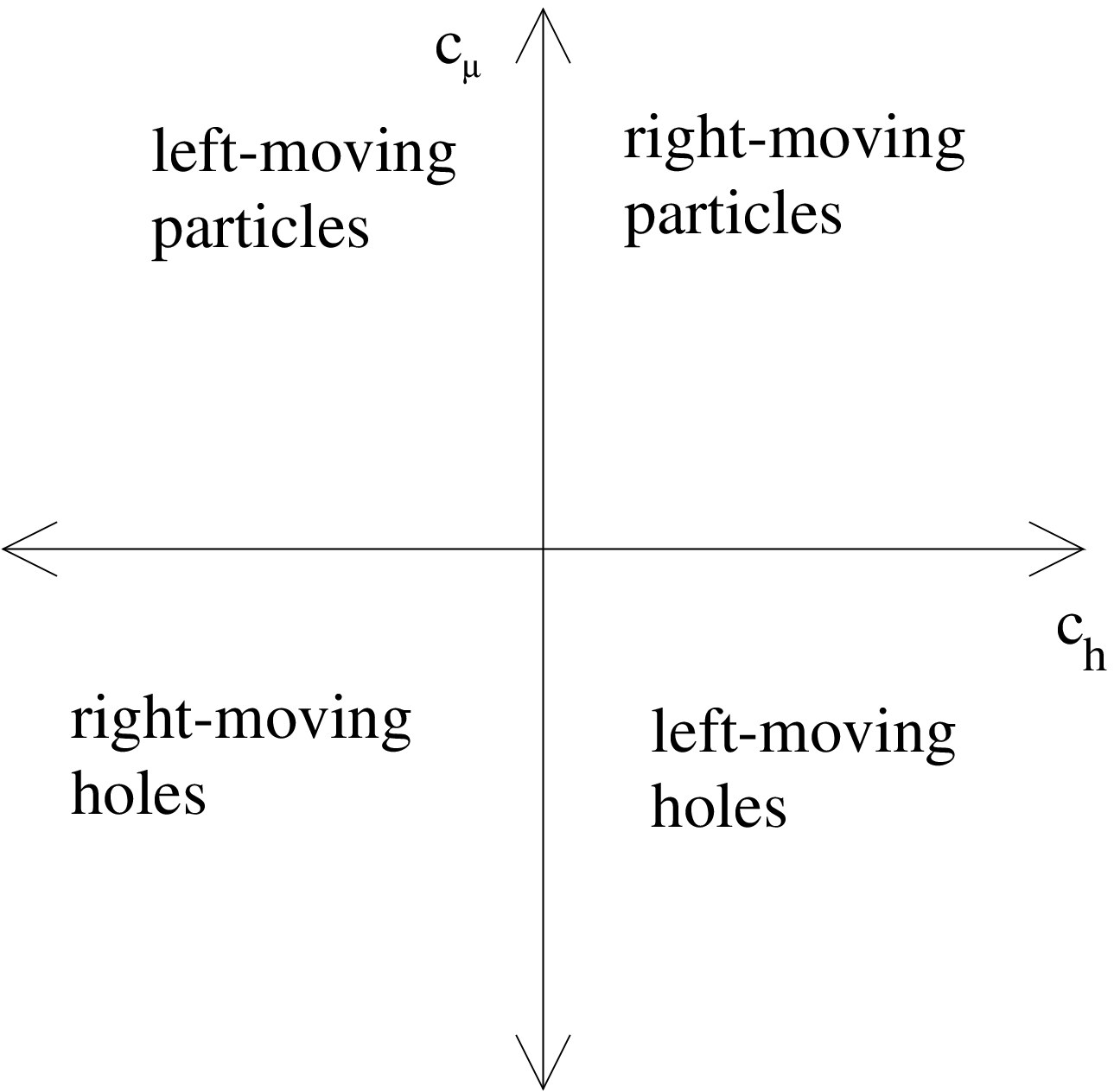}
\begin{small}
FIG.\ 5.  
The range of parameter space in which we expect to see various kinds
of defects as we cross to the superfluid side of the Mott
insulator/superfluid transition.  The parameters $c_{\mu}$ and ${\bf
c}_{\bf h}$ are defined in Eqs.\ (\ref{eq:cmu}) and (\ref{eq:ch}).
\end{small}
\vspace{0.2in}
\end{minipage}

\subsection{Relationship between defect angle and the direction of the
applied magnetic field}

Note that these angles are not the same as the direction of the excess
applied field, defined to be the additional field beyond that which
produces the Mott insulator background.  We denote this excess field
by $\delta {\bf H} \equiv {\bf H} - H_0 \hat{\bf z}$.  Because there
is actually a \textit{range} of fields that produce the Mott insulator
phase (see Fig.\ 2), we define $H_0$ so that $c_{\mu} = 0$ when
$\delta H_z = 0$.  Then a comparison of the direction of the defects
with the direction of $\delta {\bf H}$ near the transition $r=0$ gives
\begin{eqnarray}
\tan \phi_v & = & \frac{\sinh h_y}{\sinh h_x}, \\
\tan \phi_{\bf H} & = & \frac{h_y}{h_x}
\end{eqnarray}
for the azimuthal angles, and
\begin{eqnarray}
\tan \theta_v & = & \frac{a_0 t (U + \mu)}{k_B T \Delta \mu_0}
\sqrt{\sinh^2 h_x + \sinh^2 h_y}, \\
\tan \theta_{\bf H} & = & \frac{k_B T|{\bf h}|}{a_0 \Delta \mu_0}
\end{eqnarray}
for the polar angles, where
\begin{eqnarray}
\label{eq:dmu0}
\Delta \mu_0 & = & \frac{1}{2} [E_h - E_p + 2t(\cosh h_x + \cosh h_y)] \\
& \equiv & \mu - \mu_0,
\end{eqnarray}
with
\begin{equation}
\mu_0 = U (n - 1/2) - t (\cosh h_x + \cosh h_y).
\label{eq:mu0}
\end{equation}
From the equations for the azimuthal angles, we can see that the
defects will tend to penetrate along the crystallographic axes defined
by the square lattice of columnar defects.  This is because the kink
energy is smaller for hopping along the nearest neighbor directions,
and is further emphasized by Fig.\ 6, which shows the relationship
between $\phi_v$ and $\phi_{\bf H}$.  To better compare $\theta_v$ and
$\theta_{\bf H}$, we note that at the transition, we can write the
ratio
\begin{equation}
\frac{\tan \theta_v}{\tan \theta_{\bf H}} = \frac{a_0^2 t U}{(k_B
T)^2} f(\mu, {\bf h}),
\end{equation}
where
\begin{equation}
f(\mu, {\bf h}) = \left( 1 + \frac{\mu}{U} \right) \frac{\sqrt{\sinh^2
h_x + \sinh^2 h_y}}{| {\bf h}|},
\end{equation}
and $\mu$ and ${\bf h}$ are related by the constraint that we are at
the transition $r = 0$.  The prefactor $a_0^2 t U / (k_B T)^2$ gives
the order of magnitude of the ratio of $\tan \theta_v$ to $\tan
\theta_{\bf H}$, as the (dimensionless) function $f(\mu, {\bf h})$ is
of order 1 unless $|{\bf h}| \gg 1$.

\begin{minipage}[t]{3.2in}
\vspace{0.1in}
\epsfxsize=3.2in
\epsfbox{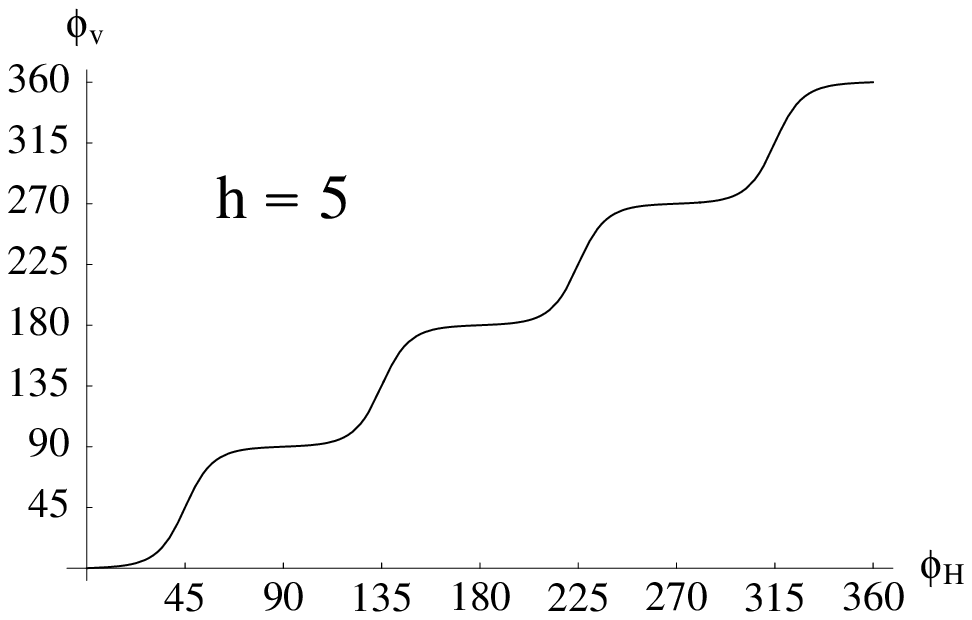}
\begin{small}
FIG.\ 6.
A plot of the azimuthal angle of the average direction of defects
versus the azimuthal angle of the excess applied field, for a fixed
$|{\bf H}_{\perp}|$.
\end{small}
\vspace{0.2in}
\end{minipage}

\subsection{Scaling analysis near the multicritical point}

A formal scaling analysis determines the shape of the transition line
near the multicritical \textit{XY} transition.  Generalized to ${\bf
c_h} \neq 0$, the scaling form~\cite{Fisher} for the singular part of
the free energy $f_s$ near the multicritical point $c = {\bf c}_{\bf
h} = 0, \delta \equiv r - r_c = 0$ reads
\begin{equation}
f_s (\delta,{\bf c}_{\bf h}, c) \sim b^{-d} f_s (\delta
b^{1/\nu_{XY}}, {\bf c}_{\bf h}b^{\lambda_{\bf h}}, c_{\mu}
b^{\lambda_{\mu}})
\label{eq:scaling}
\end{equation}
for a length rescaling parameter $b$.  $\lambda_{\bf h}$ and
$\lambda_{\mu}$ are the renormalization group eigenvalues for ${\bf
c}_{\bf h}$ and $c_{\mu}$ respectively.  Although the scaling
prefactor is $b^{-(d-1) -z}$ in general, we use the fact that $z=1$
near the \textit{XY} transition.

If we choose $b$ such that $|\delta| b^{1/\nu_{XY}} = 1$, we obtain
\begin{equation}
f_s (\delta, {\bf c}_{\bf h}, c_{\mu}) \sim |\delta|^{d \nu_{XY}}
\Phi_{\pm} ({\bf c}_{\bf h}|\delta|^{- \nu_{XY} \lambda_{\bf h}},
c_{\mu} |\delta|^{- \nu_{XY} \lambda_{\mu}}),
\label{eq:scaling2}
\end{equation}
where $\Phi_{\pm} ({\bf x}, z)$ is a scaling function that depends on
whether $\delta > 0$ or $\delta < 0$.

A straightforward generalization of arguments in Ref.~\cite{Fisher}
yields $\lambda_{\bf h} = \lambda_{\mu} = 1$ and Eq.\
(\ref{eq:scaling2}) becomes
\begin{equation}
f_s (\delta, {\bf c}_{\bf h}, c_{\mu}) \sim |\delta|^{d \nu_{XY}}
\Phi_{\pm} ({\bf c}_{\bf h}|\delta|^{- \nu_{XY}}, c_{\mu} |\delta|^{-
\nu_{XY}}).
\label{eq:scaling3}
\end{equation}
The function $\Phi_- ({\bf x},z)$ is singular along a curve $[{\bf
x}_c (z), z]$, which is necessary to describe the superfluid/Mott
insulator transition away from the special multicritical pint at
$c_{\mu} = {\bf c_h} = 0$.  Thus ${\bf c}_{\bf h} \sim c_{\mu} \sim
|\delta |^{\nu_{XY}}$ on the transition line.  Since $\nu_{XY} <1$ in
three dimensions, there is no cusp in the phase diagram, in contrast
to the Bose Glass phase.~\cite{N+Vinokur}

\section{Mott insulator phase}
\label{sec-Mott}
\setcounter{equation}{0}

The renormalization group analysis described in Appendix~\ref{sec-rg}
allows us to relate quantities near the transition to quantities deep
within the Mott insulator or superfluid phases.  The Mott insulator is
characterized by the energy gaps for excitations.  The magnitude of
these gaps will control transport properties of the sample, as
exemplified in Sec.~\ref{sec-resist} where we calculate the nonlinear
resistivity in the Mott insulator phase.

We can study these gaps far from the transition by expanding Eq.\
(\ref{eq:S}) to lowest order in $\psi$, keeping all terms quadratic
in $\psi$, not just the most relevant in a small ${\bf k}$ and
$\omega$ expansion.  As shown in Appendix~\ref{sec-expansion},
\begin{eqnarray}
\displaystyle
S(\psi) && \approx S_0(\psi) = \beta a_0^2 \int \frac{d^2 {\bf k} d
\omega}{(2\pi)^3} |\psi({\bf k}, \omega)|^2 \nonumber \\
\displaystyle
&& \times \left[ \widetilde{J^{-1}}({\bf k}) - \frac{n}{E_h + i
\omega k_B T} - \frac{n+1}{E_p - i \omega k_B T} \right].
\label{eq:quad}
\end{eqnarray}
Here $(\widetilde{J^{-1}})({\bf k})$ is the Fourier transform of
$(J^{-1})_{ij}$, given by
\begin{equation}
\widetilde{J^{-1}}({\bf k}) = \frac{1}{2t} \frac{1}{\cos (k_x a_0 +
ih_x) + \cos (k_y a_0 + ih_y)}
\label{eq:j}
\end{equation}
for a square lattice of columnar pinning sites, and $E_p$ and $E_h$
are defined by Eqs.\ (\ref{eq:barepart}) and (\ref{eq:barehole}).
The particle energy $E_p$ is the energy arising from Eq.\
(\ref{eq:hubbard}) of a ``frozen'' extra vortex, i.e., with no
contribution from the entropy gain due to wandering of the particle.
In the quantum analogy, this entropy is represented by zero point
motion.  Similarly, $E_h$ is the hole energy associated with a missing
vortex frozen at a single site of a Mott insulator state with $n$
vortices per site.

We remark that the quadratic expansion in $\psi$ is sufficient to
determine the Mott insulator/superfluid phase boundary within mean
field theory.  The Mott insulator phase is stable provided the
coefficient of the $|\psi({\bf k}, \omega)|^2$ term is positive for
all ${\bf k}$ and $\omega$.  Thus the mean field transition is
determined by the first mode that vanishes, which is the ${\bf k} =
{\bf 0}, \omega = 0$ mode.  Therefore, the transition occurs at
$r(\mu, {\bf H}_{\perp}, T)=0$, where $r$, the parameter given in Eq.\
(\ref{eq:expandquad}), is given by
\begin{equation}
r = \frac{1}{2t} \frac{1}{\cosh h_x + \cosh h_y} - \frac{n}{E_h} -
\frac{n+1}{E_p}.
\label{eq:r}
\end{equation}
The decoupling method also leads to a transition at $r=0$, as shown in
Appendix~\ref{sec-mft}.

\subsection{Energy gaps for particles and holes}

Because quantities deep in the Mott insulator phase are well described
by Gaussian fluctuations, we focus on the quadratic approximation, and
describe later how the results change close to the Mott
insulator/superfluid transition.

\subsubsection{Gaussian fluctuation effects}

To determine the free energy cost per unit length of adding a particle
or a hole to the Mott insulator phase, we show that
\begin{eqnarray}
G(\tau) & = & \langle \psi ({\bf r}, \tau) \psi^*({\bf r}, 0)
\rangle \\
& \sim & \left\{
\begin{array}{ll}
e^{- F_p \tau / k_B T} & \mbox{for $\tau > 0$} \\
e^{- F_h | \tau |/ k_B T} & \mbox{for $\tau < 0$}
\end{array} \right.
\end{eqnarray}
in the $\tau \rightarrow \infty$ limit.  This Green's function
describes an extra vortex line created and destroyed at position
$\textbf{r}$ for a time $\tau$ when $\tau$ is positive.  For $\tau <
0$, it represents a hole.  We identify $F_p$ with the (free) energy
cost per unit length of a particle and $F_h$ with the (free) energy
cost per unit length of a hole in the Mott insulator.  These energies
will be lower than the energy per unit length of particles or holes
localized to a single site ($E_p$ and $E_h$), because they can gain
entropy from wandering and hence lower their free
energies.~\cite{Frey+N+F} In the quantum analogy, these energies are
the ``gaps'' associated with particle and hole excitations in the Mott
insulator phase.

The Hubbard-Stratanovich transformed action of Eq.\ (\ref{eq:quad})
immediately yields
\begin{equation}
G(\tau) = \int \frac{d^2 k}{(2 \pi)^2} G({\bf k}, \tau),
\end{equation}
with
\begin{eqnarray}
\displaystyle
&& G({\bf k}, \tau) \nonumber \\
\displaystyle
&& = \hspace{0.1in} \frac{1}{\beta a_0^2} \int_{- \infty}^{\infty}
\frac{d \omega}{2\pi} \frac{e^{-i \omega
\tau}}{\widetilde{J^{-1}}({\bf k}) - \frac{n}{E_h + i \omega k_B T} -
\frac{n+1}{E_p - i \omega k_B T}}.
\label{eq:eph}
\end{eqnarray}
For $r > 0$, i.e., in the Mott insulator phase, the integrand has
\textit{two} poles, one each in the upper half and lower half of the
complex-$\omega$ plane.  Thus \textit{both} particles and holes are
represented by the single complex field $\psi$.  This is to be
contrasted with the situation usually encountered in vortex physics,
e.g.\ the one considered in Ref.~\cite{N88}, where the coherent state
complex field represents only one type of defect (vortex lines).

Upon solving for the location of the poles and then carrying out the
momentum integrals by steepest descents in the limit $|\tau|
\rightarrow \infty$, we find that the defect free energies are
\begin{eqnarray}
\displaystyle
F_{p,h} && = \left\{ (\Delta \mu_1)^2 + E_p E_h \Bigg[ 4rt
\right. \nonumber \\
\displaystyle
&& \left. \left. \mbox{} + 2 \frac{\sinh^2 (h_x/2) + \sinh^2
(h_y/2)}{\cosh h_x + \cosh h_y} \right] \right\}^{1/2} \mp \Delta \mu_1,
\label{eq:energies}
\end{eqnarray}
where the upper sign applies to particles and the lower to holes, with
\begin{eqnarray}
\label{eq:dmu1}
\Delta \mu_1 & = & \frac{1}{2} (E_h - E_p + 4t) \\
& \equiv & \mu - \mu_1,
\end{eqnarray}
and
\begin{equation}
\mu_1 = U (n - 1/2) - 2t.
\label{eq:mu1}
\end{equation}

When $\mu > \mu_1$, particles are preferred over holes, while for $\mu
< \mu_1$, holes are preferred.  If we set $\mu = \mu_1$, particle/hole
symmetry is restored.  When ${\bf h} = {\bf H}_{\perp} = {\bf 0}$,
then as we approach the Mott insulator/superfluid transition ($r
\rightarrow 0^+$) along a generic path with $\mu \neq \mu_1$, the
energy of the preferred defect vanishes like $r$, while the remaining
defect energy remains finite as $r \rightarrow 0$.  When $\mu =
\mu_1$, however, both defect energies vanish simultaneously
proportional to $\sqrt{r}$.

If we restore a transverse field, i.e. take ${\bf h} \neq 0$, then
\textit{neither} the particle nor the hole free energies defined by
Eq.\ (\ref{eq:energies}) vanish as $r \rightarrow 0$.  However, as
discussed in Sec.~\ref{sec-universality} and illustrated in Fig.\ 5,
we expect that \textit{tilted} particles or holes proliferate at the
transition at a nonzero angle relative to the $\hat{\bf z}$ axis.  To
see that the corresponding defect energy vanishes at the transition,
consider
\begin{eqnarray}
\label{eq:green}
G(l, \theta, \phi) & = & \langle \psi (\textbf{r}, \tau) \psi^*
(\textbf{0}, 0) \rangle \\
& = & \int \frac{d^2 k d \omega}{(2 \pi)^3} e^{i (k_x x + k_y y -
\omega \tau)} G({\bf k}, \omega) \\
& \sim & \exp [- F ({\theta, \phi}) l / k_B T],
\end{eqnarray}
where we parametrize the transverse and $\tau$-axis separations
embodied in Eq.\ (\ref{eq:green}) as
\begin{eqnarray}
\textbf{r} & = & (l \sin \theta \cos \phi, l \sin \theta \sin \phi), \\
\tau & = & l \cos \theta,
\end{eqnarray}
and we eventually take the $l \rightarrow \infty$ limit.

To carry out this Fourier transform, we rotate to a new coordinate
frame $({\bf k}', \omega ')$, with
\begin{equation}
\omega ' = \omega \cos \theta - k_x \sin \theta \cos \phi - k_y \sin
\theta \sin \phi
\end{equation}
so that 
\begin{equation}
G(l, \theta, \phi) = \int \frac{d^2 k' d \omega '}{(2 \pi)^3} e^{- i
\omega ' l} G({\bf k}', \omega ').
\end{equation}
We note that the $\omega '$ integral will be dominated in the $l
\rightarrow \infty$ limit by the closest singularity $\omega^*$ of $G
({\bf k}, \omega)$ to the real axis in the lower half of the complex
frequency plane, and will behave as $\exp [-F({\bf k}', \theta, \phi)
l /k_B T]$, where
\begin{equation}
F({\bf k}', \theta, \phi) = i \omega^* k_B T.
\end{equation}
The ${\bf k}'$ integrals are done by steepest descents, so
\begin{equation}
G(l,\theta,\phi) \sim e^{-F({\bf k}_*',\theta,\phi) l / k_B T},
\end{equation}
where ${\bf k}_*'$ is chosen such that $\left. \partial F({\bf k}',
\theta, \phi) / \partial {\bf k}' \right |_{{\bf k}_*'} = 0$.  $F
(\theta, \phi) $ can be found numerically for any $\theta,\phi $, but
the low energy excitations can be found quite simply.  We look for
solutions which obey
\begin{equation}
F (\theta,\phi) \rightarrow 0 \mbox{ as } r \rightarrow 0^+.
\end{equation}
In other words, the lower-half plane pole of the Green's function
should vanish as we approach the transition.  But $G^{-1} ({\bf k}' =
0, \omega ' = 0) = 0$ at the transition, and indeed this is the only
mode to vanish for $r \geq 0$.  Thus we conclude that if $(\theta,
\phi)$ is chosen appropriately to provide a low energy excitation,
then ${\bf k}_*' = 0$.

The angle that satisfies this condition is therefore defined by
\begin{equation}
\left. \frac{\partial F({\bf k}', \theta, \phi)}{\partial {\bf k}'}
\right|_{{\bf k}' = 0} = 0.
\end{equation}
Hence, in a small wavelength expansion of $G^{-1}({\bf k}', \omega
')$, there must be no terms linear in ${\bf k}'$.  But notice that in
the $(\tilde{\bf k}, \tilde{\omega})$ frame of Appendix~\ref{sec-rg},
the action has no terms linear in $\tilde{\bf k}$.  Therefore,
$(\tilde{\bf k}, \tilde{\omega}) = ({\bf k}', \omega ')$, and the
angles $(\theta, \phi)$ at which there are low energy excitations are
exactly those in Eqs.\ (\ref{eq:theta}) and (\ref{eq:phi}).  As
expected, the direction in which there are low energy excitations is
the same as the average direction of the defects in the superfluid
phase.

To find the energy of these excitations, we need to find the poles of
the Green's function at ${\bf k}' = 0$ and small $\omega$.  From Eq.\
(\ref{eq:rotate}), we immediately see that the pole of interest is at
$i \omega ' = r / |{\bf v}|$, so
\begin{equation}
F(\theta, \phi) = \frac{r}{|{\bf v}|}
\end{equation}
in the $r \rightarrow 0^+$ limit.

This analysis is clearly valid away from the multicritical point ${\bf
v} = 0$.  If ${\bf v} = 0$, then $\left. \partial F({\bf k}', \theta,
\phi) / \partial {\bf k}' \right|_{{\bf k}' = 0} = 0$ for all $\theta$
and $\phi$, so there are low energy excitations at \textit{all}
angles.  It is then clear from Eq.\ (\ref{eq:expandquad}) that, at the
level of Gaussian fluctuations,
\begin{equation}
F(\theta, \phi) \sim \sqrt{r}
\end{equation}
for all $\theta$ and $\phi$ as $r \rightarrow 0^+$, with the
coefficient depending on the angle.

\subsubsection{Defect free energy at the XY transition}

While we expect the Gaussian fluctuation analysis to be valid (up to
logarithmic corrections) near the generic transition, there will be
nontrivial corrections near the \textit{XY} transition at the
multicritical point $c_{\mu} = 0$, $\textbf{c}_{\textbf{h}} =
\textbf{0}$.  Close to this special point, we find that the free
energy $F_{XY}$ of defect lines at all angles scales in the same way
with $r$ within the Gaussian fluctuation analysis
\begin{equation}
F_{XY} \sim \sqrt{r} \sim \xi_{XY}^{-1/2},
\end{equation}
where $\xi_{XY} \sim 1/r$ is the Gaussian or mean field theory
correlation length.  To incorporate the effects of thermal
fluctuations near the transition, we note that $F_{XY}$ is actually an
inverse correlation length in the imaginary time direction.  Because
the dynamic scaling exponent near the special multicritical
\textit{XY} transition is $z=1$, we expect
\begin{equation}
F_{XY} \sim (r - r_c)^{\nu_{XY}},
\label{eq:multiscale}
\end{equation}
where $\nu_{XY} \approx 2/3$ is the correlation length exponent of the
three-dimensional \textit{XY} model, and $r_c$ is the transition point
renormalized by fluctuations.  Equation (\ref{eq:multiscale}) is, of
course, the behavior predicted for the energy gap near the transition
predicted by Fisher \textit{et al}.~\cite{Fisher}

\subsubsection{Nonlinear resistivity in the Mott insulator phase}
\label{sec-resist}

We now determine the form of the nonlinear current-voltage
characteristics for a current perpendicular to the vortex lines in the
Mott insulator phase.  Although the linear resistivity vanishes,
supercurrents in the $(x,y)$-plane generate a nonzero voltage due to
thermally activated nucleation and growth of particle-hole pairs.  To
calculate this resistivity, we estimate the free energy of a
particle-hole pair of a given size in the presence of an externally
imposed current $\textbf{J}$.  We expect that on scales much larger
than $a_0$, the lowest energy configuration will be approximately
given by a loop (see Fig.\ 7).  Then, following the analysis of
Fisher, Fisher, and Huse for a conventional Meissner
phase,~\cite{Fisher+F+H} we estimate the free energy of a loop of
radius $R \gg a_0$ lying in a plane normal to the current to be
\begin{equation}
F_{\mbox{loop}} \approx 2 \pi R \overline{F} - J \frac{\phi_0}{c} \pi
R^2,
\end{equation}
where the second term arises from work done against the Lorentz force
and the first is the line free energy of the loop.  If ${\bf J}$ is in
the $(\theta_{\bf J} = \pi/2, \phi_{\bf J})$ direction, then the
average energy per unit length of the particle-hole pair is
\begin{equation}
\overline{F} = \frac{1}{\pi} \int_{0}^{\pi} \left[ F \left( \theta,
\phi_{\bf J} - \frac{\pi}{2} \right) + F \left( \theta, \phi_{\bf J} +
\frac{\pi}{2} \right) \right] d\theta,
\end{equation}
as illustrated in Fig.\ 7.

\begin{minipage}[t]{3.2in}
\vspace{0.1in}
\epsfxsize=3.2in
\epsfbox{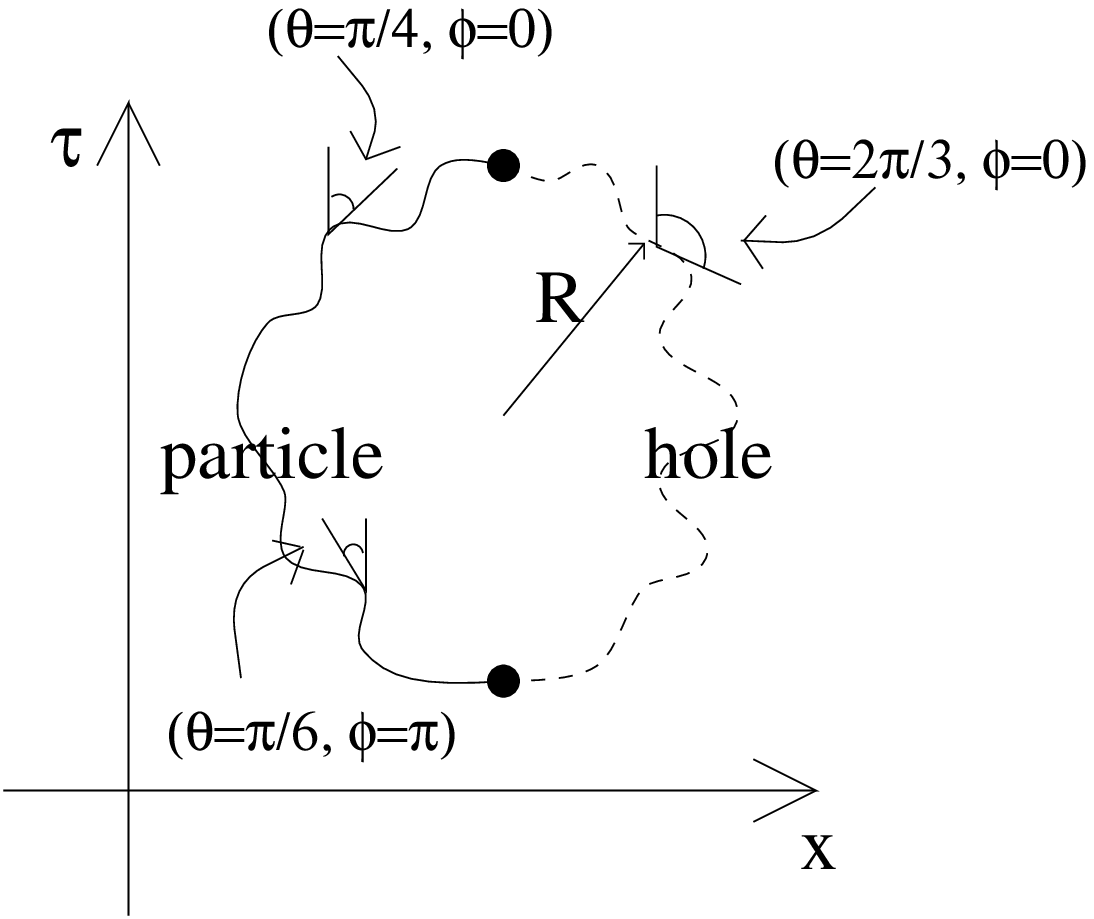}
\begin{small}
FIG.\ 7.
A particle/hole defect loop which gives rise to a nonlinear
resistivity by nucleating and growing in response to a current
perpendicular to the $(x,\tau)$ plane.
\end{small}
\vspace{0.2in}
\end{minipage}

$F_{\mbox{loop}}$ has a maximum at $R_c \approx c \overline{F} / J
\phi_0$, yielding a free energy barrier of approximately $\pi c
\overline{F}^2 / J \phi_0$.  Thus the voltage generated by loops going
over the barrier takes the form
\begin{equation}
V \sim e^{- J_T / J},
\end{equation}
with
\begin{equation}
J_T \approx \frac{\pi c \overline{F}^2}{\phi_0 T}.
\end{equation}
We expect this to be valid for
\begin{equation}
J \ll \frac{c \overline{F}}{\phi_0 a_0}.
\end{equation}

Therefore, near the \textit{XY} transition, we find from Eq.\
(\ref{eq:multiscale}) that
\begin{equation}
J_T \sim (r - r_c)^{2\nu_{XY}}.
\end{equation}
Near the generic transition, $J_T$ does not vanish, but instead
approaches a nonzero constant.  The exact value of $\overline{F}$ is
tedious to obtain, but it can be estimated as $\overline{F} \approx
\frac{1}{2} (F_p + F_h)$, or
\begin{eqnarray}
\displaystyle
\overline{F} && \approx \left\{ (\Delta \mu_1)^2 + E_p E_h \Bigg[ 4rt
\right. \nonumber \\
\displaystyle
&& \left. \left. \mbox{} + 2 \frac{\sinh^2 (h_x/2) + \sinh^2
(h_y/2)}{\cosh h_x + \cosh h_y} \right] \right\}^{1/2},
\end{eqnarray}
where $\Delta \mu_1$ is given in Eq. (\ref{eq:dmu1}).

\subsection{Finite size effects in the Mott insulator}
\label{sec-boundary}

In this section, we discuss the difference between the periodic
boundary conditions appropriate to real quantum bosons and the free
boundary conditions appropriate to flux lines.  In the Mott insulator
phase, the transverse magnetic field is screened in the bulk of the
sample where the vortices are localized on lattice sites, a phenomenon
known as the transverse Meissner effect.~\cite{N+Vinokur} However,
near the boundaries of the sample along the $\tau$-direction, the
transverse field will be able to penetrate.  We show here that surface
effects fall off exponentially with distance and calculate the
corresponding screening length.  This length scale plays a role
similar to the usual London penetration depth in the Meissner phase,
but is physically distinct, as it arises from the existence of a
periodic pinning potential.

It is convenient to characterize surface effects by $\langle \hat{\cal
H}_1 \rangle_{\tau}$, where $\langle \hat{\cal H}_1 \rangle_{\tau}$ is
the matrix element of Eq. (\ref{eq:h1}) in the Heisenberg
representation,
\begin{equation}
\langle \hat{\cal H}_1 \rangle_{\tau} = \langle \psi_f | e^{- \beta (L
- \tau) \hat{\cal H}} \hat{\cal H}_1 e^{- \beta \tau \hat{\cal H}} |
\psi_i \rangle,
\label{eq:defH1}
\end{equation}
and $| \psi_i \rangle$ and $| \psi_f \rangle$ are initial and final
boson states appropriate to ``free'' boundary conditions.  By ``free''
boundary conditions, we mean that one integrates freely over the
positions at which vortices enter and exit the sample.  As discussed
in Ref.~\cite{N88}, this boundary condition can be represented in the
boson mapping by
\begin{equation}
| \psi_i \rangle = | \psi_f \rangle = e^{\sqrt{n_0}} |0 \rangle,
\end{equation}
where $n_0 = \sum_j \hat{a}^{\dag}_j \hat{a}_j$ and $|0 \rangle$ is
the vacuum state.

Recall $\hat{\cal H}_1$ is the hopping term, which leads to a free
energy reduction due to the wandering of the flux lines.  Because
vortex fluctuations are less constrained near a free surface, we
expect that $\hat{\cal H}_1$ exceeds its bulk value near the surface
$\tau = 0$. In Appendix~\ref{sec-finite}, we show more precisely that
\begin{equation}
\langle \hat{\cal H}_1 \rangle_{\tau} - \langle \hat{\cal H}_1
\rangle_{\tau = \infty} \sim e^{- \tau / \tau^*}
\label{eq:expdecay}
\end{equation}
as $\tau \rightarrow \infty$, with
\begin{equation}
\tau^* = \frac{k_B T}{F_p + F_h},
\label{eq:London}
\end{equation}
where $F_p$ and $F_h$ are the $\textbf{h}$-dependent gaps given by
Eq.\ (\ref{eq:energies}).  Equation (\ref{eq:London}) implies that the
influence of a free surface extends over the distance a particle-hole
pair (with free energy $F_p + F_h$) can exist in thermal equilibrium
at temperature $T$, as illustrated in Fig.\ 8.

\begin{minipage}[t]{3.2in}
\vspace{0.1in}
\epsfxsize=3.2in
\epsfbox{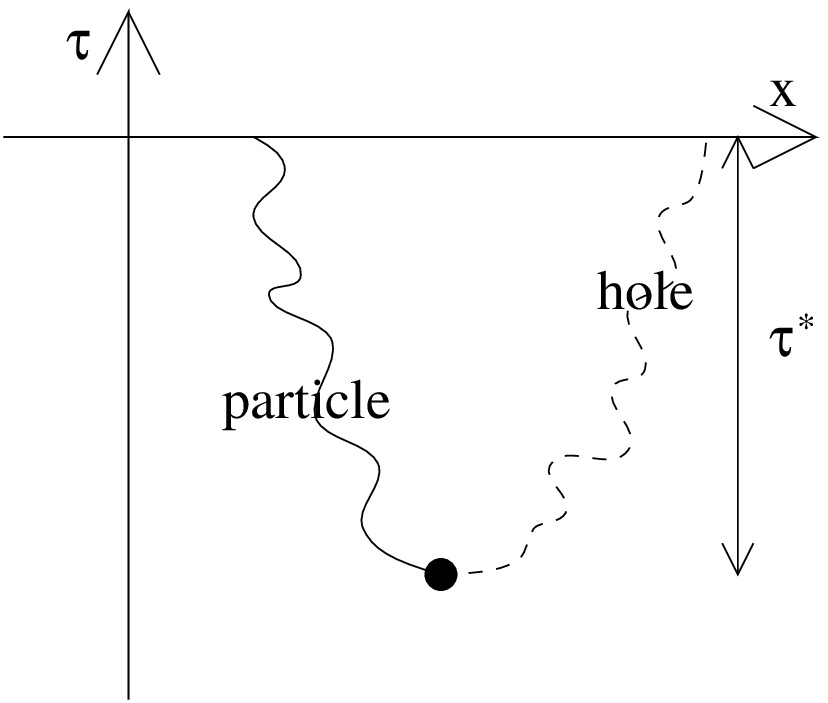}
\begin{small}
FIG.\ 8.
The screening length $\tau^*$ of a transverse magnetic field is
determined by the distance over which a particle-hole pair can exist
in thermal equilibrium.
\end{small}
\vspace{0.2in}
\end{minipage}

Away from the multicritical point, this particular ``healing length''
does \textit{not} diverge near the superfluid phase: suppose, for
example, that particles are the favored defects whose energy vanishes
as we approach the transition.  However, any particle excitation near
the boundary must be accompanied by a hole, with a much higher free
energy, yielding a finite healing length.  Once we cross the
transition into the flux liquid or ``superfluid'' phase, the free
energy per unit length of extra particles becomes negative.  Only then
can it extend across the whole length of the sample, and no longer
requires an accompanying vacancy.

For penetration at surfaces in other directions, we expect (although
we have not done explicit calculations) that the magnetic field falls
off exponentially with a decay length of
\begin{equation}
\tau^* = \frac{k_B T}{F (\theta, \phi) + F (\pi - \theta, \phi + \pi)},
\end{equation}
where $(\theta, \phi)$ is the angle of the normal to the surface.
Again, this will not diverge unless $c_{\tau} =
\textbf{c}_{\textbf{h}} = 0$.

\section{Superfluid phase}
\label{sec-super}
\setcounter{equation}{0}

To study the superfluid phase, we must include quartic powers of
$|\psi|$ in the action.  Near the transition, the truncation
(\ref{eq:expandquad}) should be an adequate approximation, because
other terms, such as $|\psi |^6$ and gradient operators combined with
$\psi^* \psi^* \psi \psi$, are irrelevant variables.

Note first that in the superfluid phase, \textit{both} particles and
holes are present in equilibrium, even away from the exceptional
multicritical point.  As illustrated in Fig.\ 9, $\langle \psi
(\textbf{r}, \tau) \rangle \sim e^{-F_+/k_B T}$, where $F_+$ is the
free energy associated with creation of a vortex ``head'' or magnetic
monopole at position $(\textbf{r}, \tau)$.  Similarly, $\langle \psi^*
(\textbf{r}, \tau) \rangle \sim e^{-F_-/k_B T}$, where $F_-$ is the
free energy associated with creation of a vortex ``tail'' or magnetic
antimonopole.~\cite{Los Alamos} In the Mott insulator phase, these
heads and tails are accompanied by long strings of (energetically
unfavorable) particle or hole excitations, so $F_+$ and $F_-$ diverge
as the sample thickness $L \rightarrow \infty$.  The presence of both
particles and holes in the superfluid phase follows because $F_+$ and
$F_-$ are finite, i.e.,
\begin{equation}
\lim_{|\tau| \rightarrow \infty} \langle \psi (\textbf{r}, \tau)
\psi^*(\textbf{r}, 0) \rangle = \langle \psi (\textbf{r}, \tau)
\rangle \langle \psi^*(\textbf{r}, 0) \rangle \neq 0.
\end{equation}
When $c \neq 0$ or $\textbf{c}_{\textbf{h}} \neq \textbf{0}$, we find
more precisely that
\begin{equation}
\lim_{|\tau| \rightarrow \infty} \langle \psi (\textbf{r}, \tau)
\psi^*(\textbf{r}, 0) \rangle = | \langle \psi \rangle |^2 \sim | r
\ln r |
\end{equation}
just above the transition.  The explicit expression for the condensate
fraction $n_0 = | \langle \psi \rangle |^2$ is given in
Appendix~\ref{sec-rg}.  Thus both particles and holes will
proliferate, even when only one of their energies vanishes at the
transition, similar to results found for vacancies and interstitials
at the supersolid transition by Frey \textit{et al}.\ for
supersolids~\cite{Frey+N+F}, and as confirmed via the decoupling
method of Appendix~\ref{sec-mft}.  The ratio of particles to holes in
the superfluid phase is more difficult to compute, and we rely on the
decoupling method of Appendix~\ref{sec-mft} for a mean field estimate
of this quantity.  More generally, we expect that the probability of
an $m$-fold particle or hole behaves as
\begin{eqnarray}
\displaystyle
\lim_{|\tau| \rightarrow \infty} && \langle [\psi (x_i, \tau)]^m
[\psi^*(x_i, 0)]^m \rangle \nonumber \\
\displaystyle
&& \sim \langle [\psi^* (x_i, \tau)] \rangle^m \langle [\psi(x_i, 0)]
\rangle^m \nonumber \\
\displaystyle
&& \sim (| r \ln r|)^m,
\end{eqnarray}
which has logarithmic corrections to the mean field results obtained
in the decoupling approximation of Appendix~\ref{sec-mft}.

\begin{minipage}[t]{3.2in}
\vspace{0.1in}
\epsfxsize=3.2in
\epsfbox{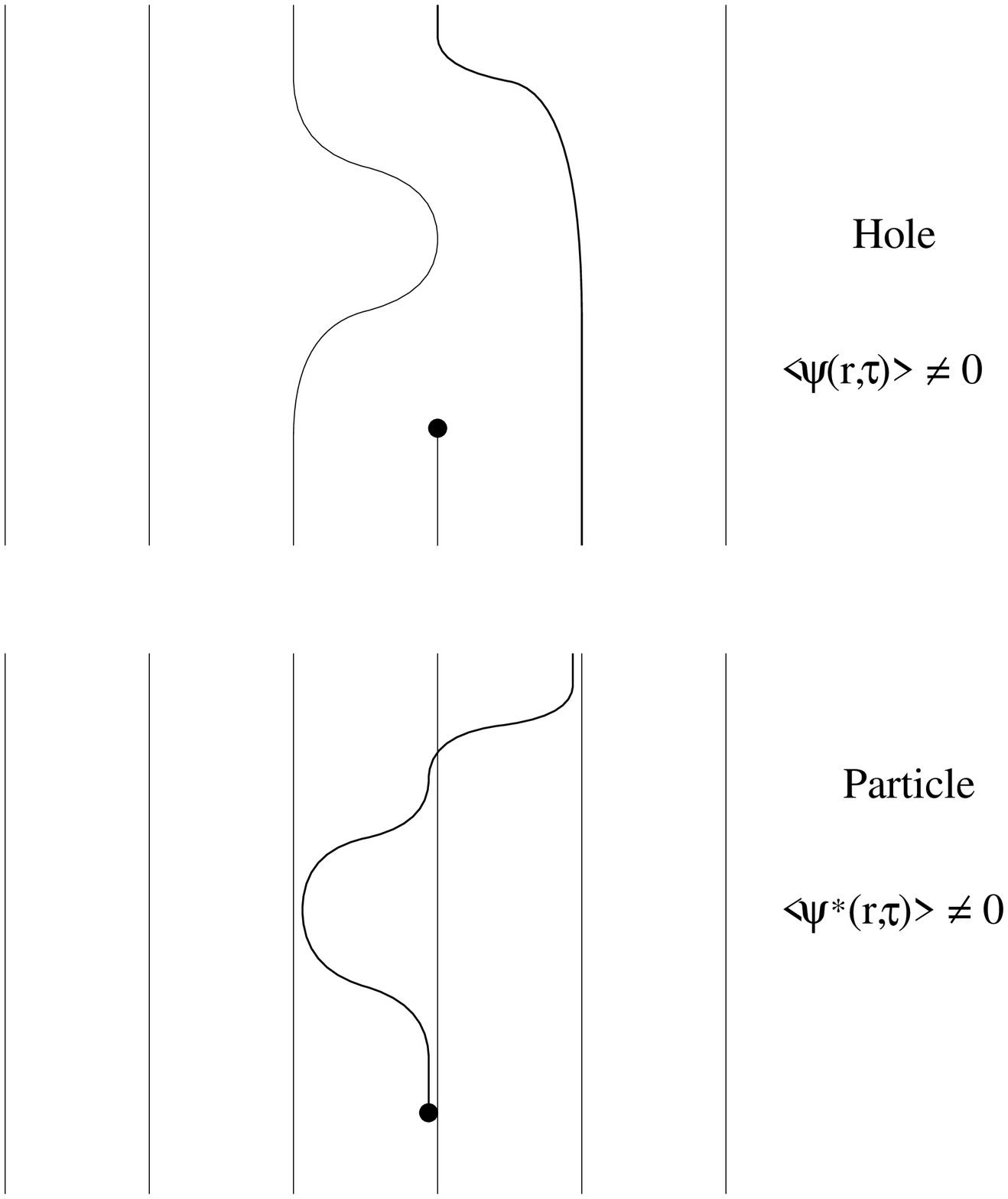}
\begin{small}
FIG.\ 9.
Vortex ``head'' and ``tail'' excitations (monopoles) associated with
$\langle \psi \rangle$ and $\langle \psi^* \rangle$.  The
corresponding free energies $F_+$ and $F_-$ are finite in the
superfluid phase, leading to $\langle \psi \rangle \neq 0$ and
$\langle \psi^* \rangle \neq 0$, while in the Mott insulator phase
there is a free energy cost \textit{per unit length}, giving rise
to $\langle \psi \rangle = \langle \psi^* \rangle = 0$ as $L_z
\rightarrow \infty$.
\end{small}
\vspace{0.2in}
\end{minipage}

\subsection{Magnetic field penetration}

Once in the superfluid phase, the magnetic field ${\bf B}$ deviates
from its frozen Mott insulator value ${\bf B}_0 = (n \phi_0 / a_0^2)
\hat{\bf z}$, unless the parameters are tuned to the multicritical
point.  Let $\delta {\bf B} = {\bf B} - {\bf B}_0$.  We show
explicitly that the excess magnetic field $\delta {\bf B}$ is
determined by the angle that which tilted particles and holes enter at
the transition.

The calculation of the magnetic field appears in
Appendix~\ref{sec-magnet}.  The excess field (to lowest order in the
condensate fraction) is given by
\begin{equation}
\delta {\bf B} = \phi_0 {\bf v} \frac{|r \ln r|}{16 \pi D}
\end{equation}
as $r \rightarrow 0$, where ${\bf v}$ is defined in Eq.\
(\ref{eq:angle}) and $D$ is given in Eq.\ (\ref{eq:D}).  This adds
further support to our claim that the transition is characterized by
the penetration of ``tilted defects,'' as the excess magnetic field is
indeed in the direction of these putative defects.  Note that the
magnitude of this excess field has the same dependence on the distance
from the Mott insulator phase as it does in the pure case near the
Meissner phase at $H_{c1}$.~\cite{N88}

\subsection{Excitation spectrum}

We now discuss the spectrum of excitations in the superfluid phase.
In the superfluid or entangled flux liquid phase, $\langle \psi ({\bf
r}, \tau) \psi^* ({\bf 0}, 0) \rangle$ exhibits long range order, as
was discussed at the beginning of this section.  Here, we discuss the
flux lines themselves, and focus on correlations in the local vortex
density
\begin{equation}
\hat{n}_i (\tau) = \hat{a}^{\dag}_i (\tau) \hat{a}_i (\tau),
\end{equation}
which are controlled by the excitation spectrum of the equivalent
quantum superfluid.~\cite{N88} Specifically, we expect that the
spatial Fourier transform $\hat{n} ({\bf k},\tau)$ of the density has
the correlation
\begin{eqnarray}
S ({\bf k}, \tau) & \equiv & \langle \hat{n}({\bf k},\tau) \hat{n}^*
({\bf k}, 0) \rangle \\
& \sim & S ({\bf k}, \tau = 0) \exp [ - \epsilon_B ({\bf k}) \tau /
k_B T]
\label{eq:structure}
\end{eqnarray}
in the $\tau \rightarrow \infty$ limit, where the decay in the
imaginary time direction is controlled by the phonon-roton spectrum
$\epsilon_B ({\bf k})$ of the superfluid.~\cite{Lieber} Our interest
here is to find the form of this spectrum when the magnetic field is
tilted to make the problem non-Hermitian.  We assume the tilt is large
enough to destroy the transverse Meissner defect and apply the
Bogoliubov approximation~\cite{Bogoliubov} to Eq.\
(\ref{eq:hubbard}), which will be accurate deep in the superfluid
phase.  This approximation treats creation and annihilation operators
in the zero momentum Fourier mode as $c$-numbers, and then neglects
interactions between bosons that are both outside the consdensate.
After some straightforward calculations, we find that the excitation
spectrum that enters Eq.\ (\ref{eq:structure}) is
\begin{equation}
\epsilon_B ({\bf k}) = \sqrt{2Un_0 \epsilon_R ({\bf k}) + \epsilon_R^2
({\bf k})} + i \epsilon_I ({\bf k}),
\label{eq:spectrum}
\end{equation}
where $n_0$ is the average vortex density,
\begin{eqnarray}
\displaystyle
\epsilon_R ({\bf k}) = 2t [ && \cosh h_x (1 - \cos k_x a_0) \nonumber \\
\displaystyle
&& \mbox{} + \cosh h_y (1 - \cos k_y a_0)],
\label{eq:real}
\end{eqnarray}
and
\begin{equation}
\epsilon_I ({\bf k}) = 2t [\sinh h_x \sin k_x a_0 + \sinh h_y \sin k_y
a_0].
\end{equation}
The real part of the spectrum (\ref{eq:real}) reduces to the usual
lattice Bogoliubov spectrum when ${\bf h} = {\bf 0}$.  The imaginary
part can be understood by considering the limit of small ${\bf k}$,
when Eq.\ (\ref{eq:spectrum}) can be written
\begin{equation}
\epsilon_B ({\bf k}) = \sqrt{2Un_0 \epsilon_R ({\bf k})} + i {\bf u_h}
\cdot {\bf k},
\end{equation}
where
\begin{equation}
{\bf u_h} = 2a_0 t (\sinh h_x, \sinh h_y)
\label{eq:Landau}
\end{equation}
is parallel to the tilt direction ${\bf c_h}$ defined in Eq.\
(\ref{eq:ch}).  Equation (\ref{eq:Landau}) is just the famous Landau
formula for excitations in a moving superfluid,~\cite{Bogoliubov}
generalized to imaginary time.  The imaginary part is required here to
describe the drift of the entangled vortex lines relative to the
column direction.  This complex spectrum is similar to the spectrum of
eigenvalues found for superfluid columnar in positions in the
non-interacting case.~\cite{Hatano+N,interest}.

\acknowledgements

This research was supported primarily by the Harvard Materials
Research Science and Engineering Laboratory through Grant
No. DMR94-00396 and by the National Science Foundation through Grant
No. DMR97-14725.  One of us (R.A.L.) acknowledges support from the
U.S. Office of Naval Research.

\appendix
\renewcommand{\theequation}{\Alph{section}\arabic{equation}}

\section{Mean field Decoupling}
\label{sec-mft}

Following Sheshadri \textit{et al}.~\cite{Sheshadri} we make the
decoupling approximation
\begin{equation}
\hat{a}_i^{\dag} \hat{a}_j \approx \langle \hat{a}_i^{\dag} \rangle
\hat{a}_j + \hat{a}_i^{\dag} \langle \hat{a}_j \rangle - \langle
\hat{a}_i^{\dag} \rangle \langle \hat{a}_j \rangle,
\end{equation}
so that the Hamiltonian (\ref{eq:hubbard}) takes the form $\hat{\cal
H} = \sum_i \hat{\cal H}_i$, with
\begin{equation}
\hat{\cal H}_i = \frac{U}{2} \hat{n}_i ( \hat{n}_i - 1) - \mu
\hat{n}_i - t^* [\psi \hat{a}_i^{\dag} + \psi^* \hat{a}_j - |\psi|^2],
\label{eq:single}
\end{equation}
and where the order parameter is $\psi = \langle \hat{a}_i \rangle$.
The effective hopping strength $t^*$ is given by
\begin{equation}
t^* = 2t (\cosh h_x + \cosh h_y).
\end{equation}
Since the sites are now decoupled in Eq.\ (\ref{eq:single}), we focus
on a single site, and drop the index $i$.  We expand the ground state
wave function of the site $| \Phi \rangle$ in the occupation number
basis:
\begin{equation}
| \Phi \rangle = \sum_{n=0}^{\infty} f(n) | n \rangle,
\end{equation}
where $\sum |f(n)|^2 = 1$ to insure proper normalization of $|\Phi
\rangle$, and $|f(n)|^2$ is the probability of finding $n$
particles on a site.  Then
\begin{eqnarray}
\label{eq:preansatz1}
\psi = \langle \hat{a}_i \rangle & = &
\displaystyle
\sum_{n=0}^{\infty} \sqrt{n+1} f^*(n) f(n+1), \\
\label{eq:preansatz2}
\psi^* = \langle \hat{a}_i^{\dag} \rangle & = &
\displaystyle
\sum_{n=0}^{\infty} \sqrt{n+1} f(n) f^*(n+1),
\end{eqnarray}
and
\begin{eqnarray}
E = \langle \hat{\cal H} \rangle = \sum_{n=0}^{\infty}
&& \left(\frac{U}{2} n(n-1) - \mu n \right) |f(n)|^2 \nonumber \\
\displaystyle
&& \mbox{} - t^* |\psi|^2.
\label{eq:preansatz3}
\end{eqnarray}
For any fixed set of magnitudes $\{|f(n)|\}$ we can maximize $|\psi|$,
and hence minimize $E$, by equating all the phases of the $f(n)$.  So
without loss of generality we assume that all $f(n)$ are real in the
ground state wave function.
 
In the Mott insulator phase with $n$ flux lines per site, we expect
$f(n') = \delta_{n,n'}$, and it follows from Eq. (\ref{eq:preansatz1})
that the order parameter $\psi = 0$.  In the superfluid phase, $f(n')$
will in general be nonzero for all nonnegative integers $n'$; however,
near the transition, we expect defects with either $n+1$ or $n-1$ flux
lines per sites to dominate.  Hence, we neglect all other types of
defects and write
\begin{equation}
f(n') = \left\{ \begin{array}{ll}
	\sqrt{\alpha} & \mbox{if $n' = n + 1$} \\
	\sqrt{\beta} & \mbox{if $n' = n - 1$} \\
	\sqrt{1 - \alpha - \beta} & \mbox{if $n' = n$} \\
	0 & \mbox{otherwise},
	\end{array}
	\right .
\label{eq:ansatz}
\end{equation}
where $\alpha$ and $\beta$ are variational parameters.  If we
reexpress $\alpha$ and $\beta$ as
\begin{eqnarray}
\alpha & = & \epsilon (1 + \omega), \\
\beta & = & \epsilon (1 - \omega),
\end{eqnarray}
then $\epsilon$ is the average number of particles \textit{and} holes
($\epsilon \ll 1$ near the transition) and $\omega$ is the
particle-hole asymmetry.  Upon expanding Eq.\ (\ref{eq:preansatz3})
to linear order in $\epsilon$, we obtain
\begin{eqnarray}
E = \epsilon [ && (E_p + E_h) + \omega (E_p - E_h) \nonumber \\
\displaystyle
&& \mbox{} - t^* (\sqrt{n+1} \sqrt{1 + \omega} + \sqrt{n} \sqrt{1 -
\omega})^2 ],
\end{eqnarray}
with $E_p$ and $E_h$ given by Eq.\ (\ref{eq:barepart}) and
Eq.\ (\ref{eq:barehole}).  By minimizing the bracketed term with respect
to $\omega$, and equating the result to zero, we find a transition
from the Mott insulator to the superfluid in agreement with the
Hubbard-Stratanovich result (\ref{eq:r}).  The particle-hole
asymmetry is given by
\begin{equation}
\omega = \frac{\mu - \mu_0}{ \sqrt{(\mu - \mu_0)^2 + n (n+1) (t^*)^2}},
\end{equation}
where
\begin{equation}
\mu_0 = U (n - 1/2) - t^*/2.
\end{equation}

We can also calculate the probability of finding more exotic defects,
e.g., \textit{two} extra particles on a site, as a check on the ansatz
(\ref{eq:ansatz}).  To this end, we minimize Eq.\
(\ref{eq:preansatz3}) with respect to $n'$, with $n' \neq n$, subject
to the constraint
\begin{equation}
f(n) = \sqrt{1 - \sum_{n' \neq n} [f(n')]^2}
\end{equation}
and obtain
\begin{equation}
f(n') = \frac{t^* \psi [f(n'+1) \sqrt{n'+1} + f(n'-1)
\sqrt{n'}]}{C_{n'} + t^* \psi \kappa}
\end{equation}
for $n' \neq n$, with
\begin{eqnarray}
\psi & = & \sum_{n'=0}^{\infty} \sqrt{n'+1} f(n') f(n'+1), \\
\kappa & = & \frac{[f(n+1) \sqrt{n+1} + f(n-1)
\sqrt{n}]}{f(n)},
\end{eqnarray}
and
\begin{equation}
C_{n'} = \frac{U}{2} (n'-n)^2 - [\mu - U (n- 1/2)](n'-n).
\end{equation}
Although these equations are difficult to solve analytically, it is
possible to extract the power law dependence of $f(n')$ on $\epsilon$
near the transition.  Because $\psi \sim \kappa \sim \sqrt{\epsilon}$,
we have a recursion relation
\begin{equation}
f(n') \sim \sqrt{\epsilon}[f(n'+1) + f(n'-1)],
\end{equation}
with asymptotic solution
\begin{equation}
f(n') \sim \epsilon ^{|n'-n|/2}
\end{equation}
as $\epsilon \rightarrow 0$.  Thus the fraction of sites with $n'$
particles per site falls off like $\sim \epsilon^{|n'-n|}$ near the
transition; the same result follows from the Hubbard-Stratanovich
method.

\section{Expansion of effective action}
\label{sec-expansion}

To expand the effective action of Eq.\ (\ref{eq:S}) in
powers of $\psi$, first define the Fourier transform
\begin{equation}
\psi ({\bf k}, \omega) = \sum_i \int_0^{L} d\tau \psi_i (\tau) e^{-i
({\bf k} \cdot {\bf x}_i - \omega \tau)},
\end{equation}
\begin{equation}
\psi^* ({\bf k}, \omega) = \sum_i \int_0^{L} d\tau \psi^*_i (\tau)
e^{i ({\bf k} \cdot {\bf x}_i - \omega \tau)},
\end{equation}
and
\begin{equation}
\widetilde{J^{-1}}({\bf k}) = \sum_i \left( J^{-1} \right)_{ij} e^{-i
{\bf k} \cdot ({\bf x}_i - {\bf x}_j)},
\end{equation}
with periodic boundary conditions in all directions.  Then, in the
thermodynamic limit, the first term of Eq.\ (\ref{eq:S}) reads
\begin{eqnarray}
\beta \sum_{i,j} \int_0^{L} && d \tau (J^{-1})_{ij} \psi_i^*(\tau)
\psi_j(\tau) \nonumber \\
\displaystyle
&& = \beta a_0^2 \int \frac{d^2 {\bf k} d\omega}{(2\pi)^3} \left| \psi
({\bf k}, \omega) \right|^2 \widetilde{J^{-1}}({\bf k}),
\label{eq:first}
\end{eqnarray}
with
\begin{equation}
\widetilde{J^{-1}}({\bf k}) = \frac{1}{2t} \frac{1}{\cos (k_x a_0 +
ih_x) + \cos (k_y a_0 + ih_y)}.
\end{equation}
We shall also need the second term of Eq.\ (\ref{eq:S}) expanded to
quadratic order in $\psi$, namely
\begin{eqnarray}
\displaystyle
-\frac{\beta^2}{2} \sum_i && \int_0^{L} d \tau_1 \int_0^{L} d \tau_2
\nonumber \\
\displaystyle
&& \times \left[ \psi_i (\tau_1) \psi_i^* (\tau_2) \left \langle
T_{\tau} a_i^{\dag} (\tau_1) a_i (\tau_2) \right \rangle_0
\right. \nonumber \\
\displaystyle
&& \left. \mbox{} + \psi_i^* (\tau_1) \psi_i (\tau_2) \left \langle
T_{\tau} a_i (\tau_1) a_i^{\dag} (\tau_2) \right \rangle_0 \right],
\label{eq:second}
\end{eqnarray}
where we have used the fact that $\hat{\cal H}_0$ conserves particle
number.

Upon changing variables so that the earlier time is $\tau$ and the
later time is $\tau + \delta \tau$, Eq.\ (\ref{eq:second}) becomes
\begin{eqnarray}
\displaystyle
-\beta^2 \sum_i && \int_0^{L} d \tau \int_0^{L/2} d \delta \tau
\nonumber \\
\displaystyle
&& \times \left[ \psi_i (\tau + \delta \tau) \psi_i^* (\tau) \left
\langle a_i^{\dag} (\tau + \delta \tau) a_i (\tau) \right \rangle_0
\right. \nonumber \\
\displaystyle
&& \left. \mbox{} + \psi_i^* (\tau + \delta \tau) \psi_i (\tau) \left
\langle a_i (\tau + \delta \tau) a_i^{\dag} (\tau) \right \rangle_0
\right].
\label{eq:second2}
\end{eqnarray}

The expectation values in Eq.\ (\ref{eq:second2}) can be evaluated
using the properties of the single site Hamiltonian (\ref{eq:h0}),
which is diagonal in the occupation number representation.  For
example,
\begin{eqnarray}
\displaystyle
&& \left \langle a_i^{\dag} (\tau + \delta \tau) a_i (\tau) \right
\rangle_0 \nonumber \\
\displaystyle
&& \hspace{0.4in} = \frac{\displaystyle \sum_{m=0}^{\infty} m e^{-
\beta L E_m} e^{- \beta \delta \tau (E_{m-1} - E_m)}}{\displaystyle
\sum_{m=0}^{\infty} e^{- \beta L E_m}},
\end{eqnarray}
where
\begin{equation}
E_m = \frac{U}{2} m (m-1) - \mu m.
\end{equation}
In the limit $L \rightarrow \infty$, the sums are dominated by $m=n$
such that
\begin{equation}
(n-1) < \frac{\mu}{U} < n,
\end{equation}
and we have
\begin{eqnarray}
\left \langle a_i^{\dag} (\tau + \delta \tau) a_i (\tau) \right
\rangle_0 & = & n e^{- \beta \delta \tau E_h},\\
\left \langle a_i (\tau + \delta \tau) a_i^{\dag} (\tau) \right
\rangle_0 & = & (n+1) e^{- \beta \delta \tau E_p},
\end{eqnarray}
with $E_h$ and $E_p$ the hole and particle energies given by Eqs.\
(\ref{eq:barehole}) and (\ref{eq:barepart}).

We can now Fourier transform (\ref{eq:second2}), which upon
combination with Eq.\ (\ref{eq:first}) yields the action $S (\psi)$
that appears in Eq.\ (\ref{eq:Z2}) to second order in $\psi$:
\begin{eqnarray}
\displaystyle
S(\psi) && \approx S_0(\psi) = \beta a_0^2 \int \frac{d^2 {\bf k} d
\omega}{(2\pi)^3} |\psi({\bf k}, \omega)|^2 \nonumber \\
\displaystyle
&& \times \left[ \widetilde{J^{-1}}({\bf k}) - \frac{n}{E_h + i \omega
k_B T} - \frac{n+1}{E_p - i \omega k_B T} \right].
\label{eq:appquad}
\end{eqnarray}

This quadratic approximation to the action is used to study the Mott
insulator phase in Sec.~\ref{sec-Mott}.  To determine the universality
class of the Mott insulator/superfluid transition, we need the
long-wavelength contributions (up to quadratic order in ${\bf k}$ and
$\omega$) to the coefficient of $|\psi({\bf k}, \omega)|^2$ and the
zero-frequency part of $|\psi|^4$ in the expansion of Eq.\
(\ref{eq:S}).  It is tedious, but straightforward, to find the result
used in Sec.~\ref{sec-transition}, namely
\begin{eqnarray}
\displaystyle
S(\psi) \approx && \beta a_0^2 \int \frac{d^2 {\bf k} d
\omega}{(2\pi)^3} |\psi({\bf k}, \omega)|^2 [ r - i c_{\mu} \omega + i
{\bf c}_{\bf h} \cdot {\bf k} \nonumber \\
\displaystyle
& & \mbox{} + c' \omega^2 + D_{ij} k_i k_j ] + u \int d^2 {\bf r} d
\tau |\psi({\bf r}, \tau)|^4,
\label{eq:appexpandquad}
\end{eqnarray}
where
\begin{equation}
r = \frac{1}{2t} \frac{1}{\cosh h_x + \cosh h_y} - \frac{n}{E_h} -
\frac{n+1}{E_p},
\label{eq:appr}
\end{equation}
\begin{equation}
c_{\mu} = k_B T \left( \frac{n+1}{E_p^2} - \frac{n}{E_h^2} \right),
\end{equation}
\begin{equation}
{\bf c}_{\bf h} = \frac{a_0}{2t} \frac{\hat{\bf x} \sinh h_x +
\hat{\bf y} \sinh h_y}{[\cosh h_x + \cosh h_y]^2},
\end{equation}
and
\begin{equation}
c' = (k_B T)^2 \left( \frac{n+1}{E_p^3} + \frac{n}{E_h^3} \right).
\end{equation}
The matrix $D_{ij}$ reads
\begin{equation}
D = \left(
\begin{array}{cc}
D_{xx} & D_{xy} \\
D_{xy} & D_{yy} \\
\end{array}
\right),
\end{equation}
with
\begin{eqnarray}
D_{xx} & = &\frac{a_0^2}{2t} \frac{1 + \cosh h_x \cosh h_y - \sinh^2
h_x}{(\cosh h_x + \cosh h_y)^3}, \\
D_{yy} & = &\frac{a_0^2}{2t} \frac{1 + \cosh h_x \cosh h_y - \sinh^2
h_y}{(\cosh h_x + \cosh h_y)^3}, \\
D_{x} & = &\frac{a_0^2}{2t} \frac{- 2 \sinh h_x \sinh h_y}{(\cosh h_x
+ \cosh h_y)^3},
\end{eqnarray}
while
\begin{eqnarray}
u = \beta a_0^6 && \left[ \frac{(n+1)^2}{E_p^3} + \frac{n^2}{E_h^3} +
\frac{n(n+1)(E_p + E_h)}{E_p^2 E_h^2}  \right.
\nonumber \\
&& \left. \mbox{} - \frac{(n+1)(n+2)}{E_p^2 E_{2p}} -
\frac{n(n-1)}{E_h^2 E_{2h}} \right],
\end{eqnarray}
where $E_{2p}$ and $E_{2h}$ are the energies of ``double particle''
and ``double hole'' excitations respectively,
\begin{eqnarray}
E_{2h} & = & 2 \mu - (2n-3) U,\\
E_{2p} & = & (2n+1) U - 2 \mu.
\end{eqnarray}
Note that $u > 0$ because $E_{2h} > 2E_h > 0$ and $E_{2p} > 2E_p > 0$.

\section{Renormalization group}
\label{sec-rg}

We first review the renormalization group treatment of the action of
Eq.\ (\ref{eq:expandquad}) away from the multicritical point $c =
{\bf c}_{\bf h} = 0$.  The analysis simplifies if we transform the
axes of space and time so that the imaginary time axis is no longer
the $\hat{\bf z}$ axis but coincides with the preferred direction of
the particle or hole defects.  We denote quantities in this rotated
frame with a tilde.  In this special coordinate frame, the term linear
in $\tilde{\bf k}$ vanishes and the action takes the form
\begin{eqnarray}
S(\psi) & \approx & \beta a_0^2 \int_{\tilde{\bf k}, \tilde{\omega}}
|\psi(\tilde{\bf k}, \tilde{\omega})|^2 [ r - i C \tilde{\omega} + D
\tilde{k}^2] \nonumber \\
& \mbox{} + & u \int_{\tilde{\bf r}, \tilde{\tau}} |\psi(\tilde{\bf
r}, \tilde{\tau})|^4,
\label{eq:rotate}
\end{eqnarray}
where $C = |{\bf v}|$ from Eq.\ (\ref{eq:angle}).  In general the
quadratic wave vector dependencies will take the form $D_{ij}
\tilde{k}_i \tilde{k}_j$.  However, we can rotate the spatial axes so
that the matrix $D_{ij}$ is diagonal, and then rescale the $\tilde{x}$
and $\tilde{y}$ axes (holding the area fixed) so that $D_{ij} = D
\delta_{ij}$, giving the simpler expression above, with
\begin{eqnarray}
D && = \frac{a_0^2}{2t |{\bf v}| (\cosh h_x + \cosh h_y)^3} \nonumber
\\
&& \times \left\{ \frac{c'}{2t} (\cosh h_x + \cosh h_y)(\cosh h_x
\cosh h_y - 1) \right. \nonumber \\
& & \hspace{0.1in} \mbox{} - c_{\mu}^2 \Bigg[ (\cosh^2 h_x + \cosh^2
h_y  + 2)(\cosh h_x \cosh h_y - 2) \nonumber \\
&& \hspace{1in} \mbox{} + 2 \Bigg] \Bigg\}^{1/2}. 
\label{eq:D}
\end{eqnarray}
(It can be shown that the quantity in braces is always positive when
$r \geq 0$.)  Also, there should appear an $\tilde{\omega}^2$ piece
and an $\tilde{\omega} \tilde{\bf k}$ piece.  However, these terms can
easily be shown to be irrelevant variables near the transition,
similar to possible $\tilde{\bf k}$ and $\tilde{\omega}$-dependent
terms in the quadratic coupling.  We add a high momentum cutoff, and
thus allow only $| \tilde{\bf k} | < \Lambda$.  Following, e.g.,
Ref.~\cite{Fisher+Hohenberg}, we will integrate out degrees of freedom
$\psi (\tilde{\bf k}, \tilde{\omega})$ with momenta in a shell
$\Lambda b^{-1} < | \tilde{\bf k} | < \Lambda$ (and any
$\tilde{\omega}$), and rescale positions by $b$ and imaginary times by
$b^z$ to restore our original cutoff.  We also rescale our fields by a
factor $b^{\zeta}$.  In $d = 2$, $u$ is marginally irrelevant at the
Gaussian fixed point $u = 0$, allowing us to work perturbatively in
$u$ to lowest order.  The most important diagrams for renormalizing
$r$ and $u$ are shown in Fig.\ 10.  In $d+1$ dimensions, we set $z =
2$ and $\zeta = - d/2$ to keep $C$ and $D$ fixed.  Upon letting $b =
e^l$ with $l$ infinitesimal, we obtain the renormalization group
recursion relations in $d = 2$
\begin{eqnarray}
\label{eq:rgr}
\frac{dr}{dl} & = & 2 r, \\
\label{eq:rgu}
\frac{du}{dl} & = & - \frac{u^2}{4 \pi C D \beta^2 a_0^4}.
\end{eqnarray}
Although $u$ is irrelevant at the Gaussian fixed point, its slow decay
to zero leads to logarithmic corrections to results calculated in the
mean field approximation.

\begin{minipage}[t]{3.2in}
\vspace{0.1in}
\epsfxsize=3.2in
\epsfbox{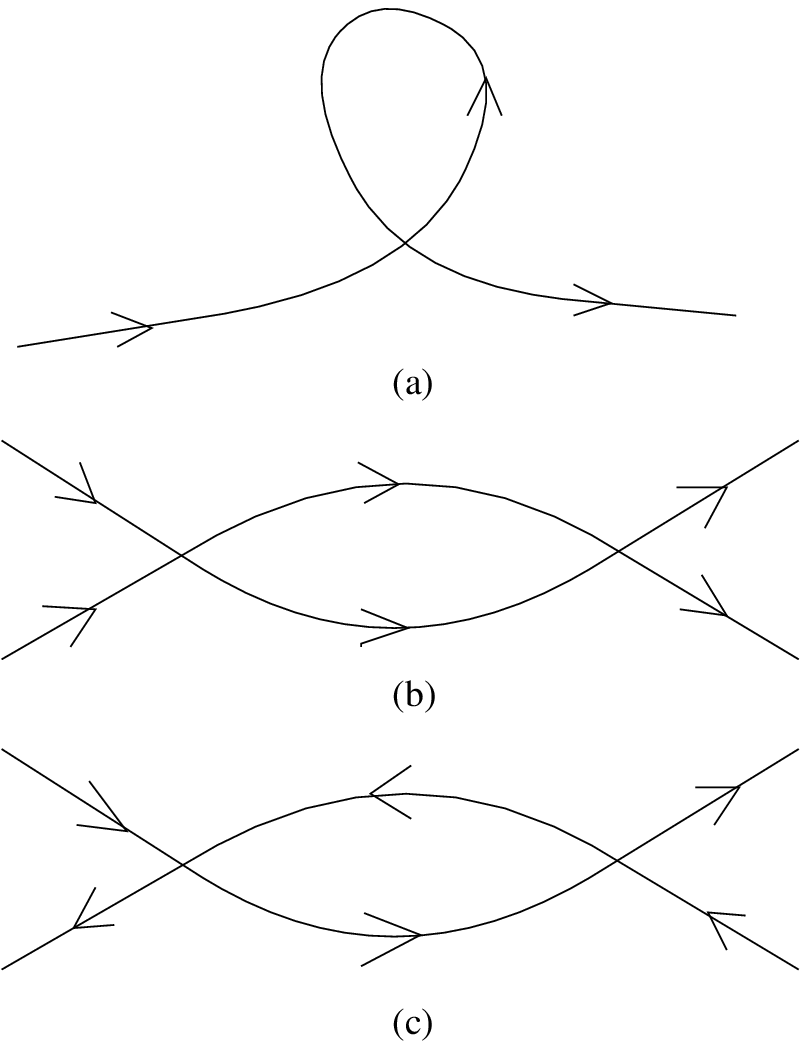}
\begin{small}
FIG.\ 10.
(a) The most important diagram renormalizing $r$.  (b) The most
important diagram renormalizing $u$.  (c) A diagram that vanishes
identically.
\end{small}
\vspace{0.2in}
\end{minipage}

To exhibit these logarithmic corrections, we first solve the recursion
relation for $u$ to get [with $u_0 = u (l=0)$]
\begin{equation}
u(l) = \frac{u_0}{1 + u_0 l / (4 \pi C D \beta^2 a_0^4)},
\end{equation}
which behaves for large $l$ like
\begin{equation}
u(l) \rightarrow \frac{4 \pi C D \beta^2 a_0^4}{l}.
\end{equation}
Consider now the condensate fraction
\begin{equation}
n_0 = | \langle \psi ({\bf r}, \tau) \rangle |^2
\end{equation}
right above the transition.  We express $n_0$ as a function of the
parameters in the Hamiltonian, and iterate them under the
renormalization group.  Then
\begin{equation}
n_0 (r, C, D, u, \ldots) \approx e^{2 l \zeta} n_0 {\bf (}r(l), C(l), D(l),
u(l), \ldots {\bf )}.
\end{equation}
We iterate until $r(l) \approx re^{2l} = 1$, i.e. $l = 1/2 | \ln r|$,
so that we are deep in the flux liquid or superfluid phase.  Since
$n_0 \approx \beta a_0^2 r(l)/2u(l)$ far from the transition, we find
using $r(l) \approx re^{2l}, u(l) \approx 4 \pi CD \beta^2 a_0^4 /l$ that
\begin{equation}
| \langle \psi \rangle |^2 = n_0 = \frac{1}{16 \pi C D \beta a_0^2} |
r \ln r|,
\end{equation}
which has logarithmic corrections from the naive result $n_0 \sim r$.

We now turn to the case where $c_{\mu} = {\bf c}_{\bf h} = 0$, so we
are at the multicritical point.  Up to a simple rescaling of lengths,
the action (\ref{eq:expandquad}) is then isotropic in space and time,
the dynamical exponent is $z=1$, and the universality class is that of
the three-dimensional \textit{XY} model.  As noted by
Josephson~\cite{Josephson}, the superfluid density scales differently
from the condensate fraction, with the condensate fraction behaving as
\begin{equation}
n_0 \sim (r_c - r)^{2 \beta}
\end{equation}
near the transition while the superfluid density behaves as
\begin{equation}
\rho_s \sim (r_c - r)^{2 \beta - \nu \eta} \sim (r_c - r)^{(d - 2)
\nu},
\end{equation}
where $r_c - r$ is the distance from the transition.  (The superfluid
density is proportional to the inverse tilt modulus in the flux line
system.~\cite{N+Vinokur,Tauber+N})  We will have
anomalous scaling for the superfluid density at the generic points of
the phase boundary as well.  Note that $\eta = 0$ since we are at the
upper critical dimension of the generic $H_{c1}$ transition, so the
corrections are limited to the logarithmic terms.

\subsection{Validity of the continuum limit}

To explore the validity of taking the continuum limit, we add a new
term into the action of Eq.\ (\ref{eq:expandquad}) that represents a
vestige of the periodic columnar array, namely
\begin{equation}
\delta S(\psi) = \beta \int d^2 {\bf x} d \tau \delta r ({\bf x})
|\psi({\bf x}, \tau)|^2,
\label{eq:periodic}
\end{equation}
where
\begin{equation}
\delta r ({\bf x}) = p \left[ \cos \left( \frac{2 \pi x}{a_0} \right) +
\cos \left( \frac{2 \pi y}{a_0} \right) \right].
\end{equation}
We then investigate how the addition of this term changes the
renormalization group.

First we consider the case ${\bf h} = {\bf 0}$, so that we do not have
to rotate our coordinate system as in Eq.\ (\ref{eq:rotate}).  Then,
in momentum space, the additional terms all involve terms such as
$\psi({\bf k}, \omega) \psi^* ({\bf k} + {\bf G}, \omega)$, where
${\bf G}$ is a reciprocal lattice vector of the underlying square
lattice.  When we reach the point in our renormalization group
procedure where $(2 \pi / a_0) e^l = \Lambda$, all of the terms
added in Eq.\ (\ref{eq:periodic}) will be integrated out, leading
only to finite changes in the remaining coupling constants.  Hence,
the critical phenomena at the Mott insulator to superfluid transition
will be unchanged by the addition of the periodic potential.  Note
that this argument applies both to the multicritical point and to the
generic transition with ${\bf h} = {\bf 0}$.

When ${\bf h} \neq {\bf 0}$, we must rotate our coordinate system as
in Eq.\ (\ref{eq:rotate}).  For most values of $c_{\mu}$ and ${\bf
c_h}$, all of the terms in Eq.\ (\ref{eq:periodic}) will still
involve a jump in the momentum $\tilde{\bf k}$, so the argument above
holds here as well.  The exception is when the direction of the
defects is perpendicular to the columns, i.e., $c_{\mu} = 0$.  Then
there will be terms in Eq.\ (\ref{eq:rotate}) that involve a jump
only in $\tilde{\omega}$, not in $\tilde{\bf k}$.  We can rewrite
these terms as
\begin{eqnarray}
\delta S(\psi) = \beta && a_0^2 \frac{p}{2} \int \frac{d^2 \tilde{\bf
k} d \tilde{\omega}}{(2\pi)^3} \psi(\tilde{\bf k}, \tilde{\omega})
\nonumber \\
&& \times \left[ \psi^* (\tilde{\bf k}, \tilde{\omega} + \Delta) +
\psi^* (\tilde{\bf k}, \tilde{\omega} - \Delta) \right],
\label{eq:prenorm}
\end{eqnarray}
where the initial value of $\Delta$ is $2 \pi / a_0$.

We now investigate how the parameters $p$ and $\Delta$ behave under
the renormalization group.  The diagrams renormalizing $r$, $p$, and
$u$ to one loop order are shown in Fig.\ 11.  Note that all diagrams
contributing to the renormalization of $p$ in this approximation
vanish identically, as do all of the extra diagrams that renormalize
$r$.  This vanishing persists to all orders in perturbation theory,
because any diagram that contains a loop (of more than one propagator)
with all the arrows pointing in the same direction, as illustrated in
Figs.\ 10(c), 11(a), and 11(b), must vanish identically.  Moreover, all
diagrams renormalizing a propagator that contain any $p$-vertex must
contain at least one such loop.  Therefore, the zero-loop recursion
relation for $p$ is accurate to all orders in perturbation theory,
yielding:
\begin{eqnarray}
\label{eq:rgp}
\frac{dp}{dl} & = & 2 p,\\
\label{eq:rgd}
\frac{d \Delta}{dl} & = & 2 \Delta.
\end{eqnarray}
The recursion relation for $r$ remains as given by Eq.\
(\ref{eq:rgr}), accurate to one loop order.  The recursion relation
for $u$, however, has additional nonvanishing terms at one loop order,
proportional to $u^2 (p / \Delta)^{2n}$, for $n$ a positive integer.
(There are also terms with extra powers of $\Delta$ in the
denominator, but those vanish in the large $l$ limit.)  Since $p /
\Delta$ is independent of $l$, only the constant in Eq.\
(\ref{eq:rgu}) is changed; the structure of the equations remains the
same.  In other words, we expect that, aside from prefactors, the
results of our renormalization group calculations will be unchanged by
the addition of a periodic potential.  Therefore we can safely neglect
terms with the periodicity of the columnar pins in our continuum
description of the Mott insulator to superfluid transition.

\begin{minipage}[t]{3.2in}
\vspace{0.1in}
\epsfxsize=3.2in
\epsfbox{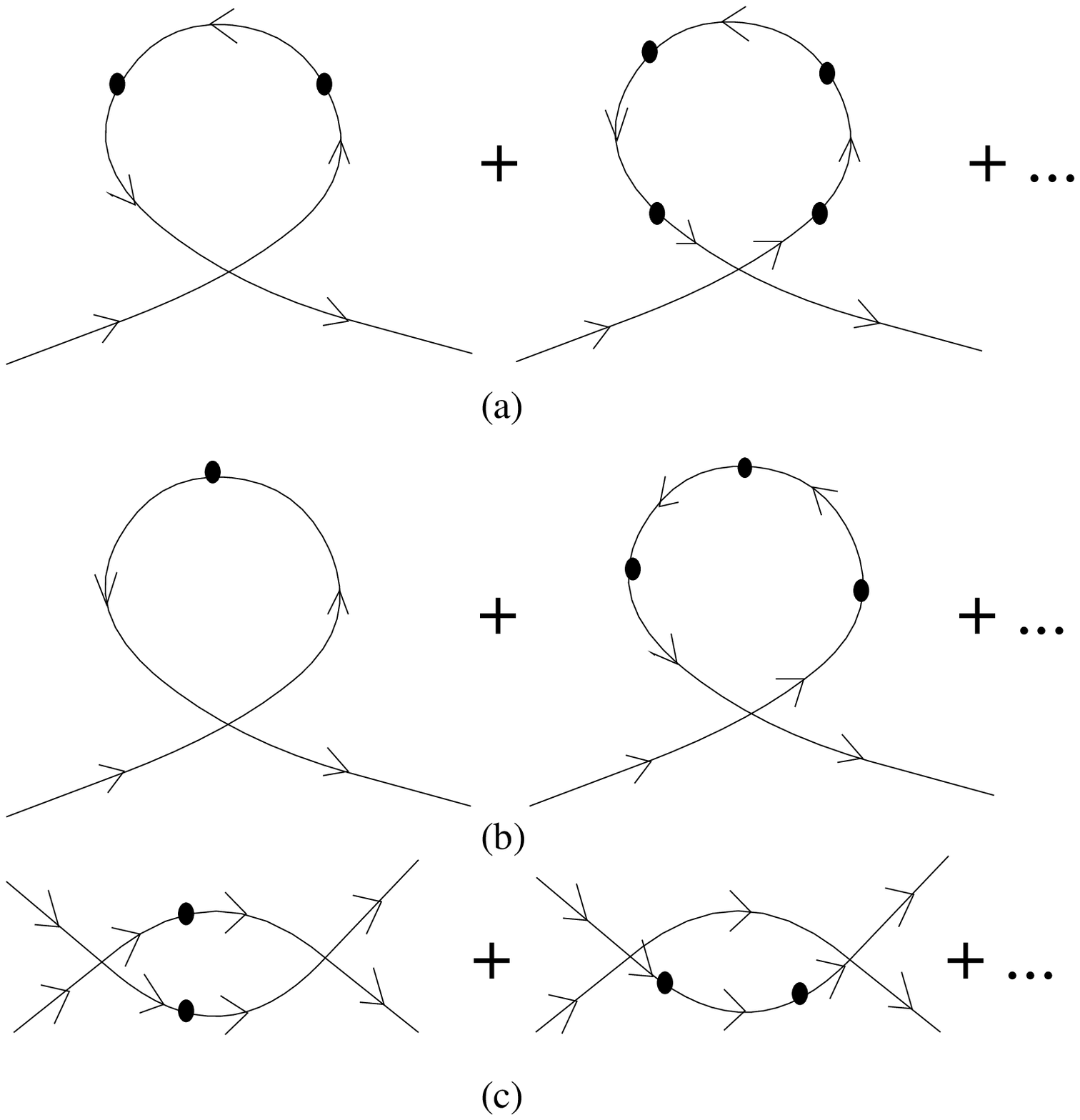}
\begin{small}
FIG.\ 11.
(a) The diagrams involving factors of $p$ that renormalize $r$.  The
dots are ``interaction'' vertices, with the interaction as given in
Eq.\ (\ref{eq:prenorm}).  The diagrams all vanish identically.  (b)
The diagrams renormalizing $p$.  These also vanish identically.  (c)
Nontrivial diagrams involving factors of $p$ that renormalize $u$.
\end{small}
\vspace{0.2in}
\end{minipage}

\section{Finite size effect}
\label{sec-finite}

In this appendix we examine a particular type of free surface effect,
by showing that
\begin{equation}
\langle \hat{\cal H}_1 \rangle_{\tau} - \langle \hat{\cal H}_1
\rangle_{\tau = \infty} \sim e^{- \tau / \tau^*},
\label{eq:H1}
\end{equation}
where $\langle \hat{\cal H}_1 \rangle_{\tau}$ is defined in Eq.\
(\ref{eq:defH1}).

In the interaction representation, Eq.\ (\ref{eq:defH1}) takes the
form
\begin{eqnarray}
\langle \hat{\cal H}_1 \rangle_{\tau} = \frac{{\cal Z}_0}{{\cal Z}}
&& \left \langle T_{\tau} \exp \left[ - \beta \int_{\tau}^{L} d \tau '
\hat{\cal H}_1 (\tau ') \right] \hat{\cal H}_1 (\tau)
\right. \nonumber \\ 
&& \times \left. T_{\tau} \exp \left[ - \beta \int_0^{\tau} d \tau '
\hat{\cal H}_1 (\tau ') \right] \right \rangle_0,
\end{eqnarray}
where
\begin{eqnarray}
{\cal Z}_0 & = & \left \langle \psi_f \left| e^{- \beta L \hat{\cal
H}_0} \right| \psi_i \right \rangle, \\
\langle \bullet \rangle_0 & = & \left \langle \psi_f \left| \bullet
e^{- \beta L \hat{\cal H}_0} \right| \psi_i \right \rangle,
\end{eqnarray}
and
\begin{equation}
{\cal Z} = \left \langle T_{\tau} \exp \left[ - \beta \int_0^{L} d
\tau ' \hat{\cal H}_1 (\tau ') \right] \right \rangle_0.
\end{equation}
If we now define a composite partition function
\begin{eqnarray}
{\cal Z}(\tau_1, \tau_2) = && \left \langle T_{\tau} \exp \left[ -
\beta \int_{\tau_2}^{L} d \tau ' \hat{\cal H}_1 (\tau ') \right]
\right. \nonumber \\
&& \times \left. T_{\tau} \exp \left[ - \beta \int_0^{\tau_1} d \tau '
\hat{\cal H}_1 (\tau ') \right] \right \rangle_0,
\end{eqnarray}
then
\begin{equation}
\langle \hat{\cal H}_1 \rangle_{\tau} = - \frac{1}{\beta}
\left. \frac{\partial}{\partial \tau} \ln {\cal Z} (\tau, \tau_2)
\right|_{\tau = \tau_2}.
\label{eq:trick}
\end{equation}
The Hubbard-Stratanovich transformation generalizes straightforwardly
to ${\cal Z}(\tau_1, \tau_2)$, and leads via Eq.\ (\ref{eq:trick}) to
\begin{equation}
\langle \hat{\cal H}_1 \rangle_{\tau} = \langle K(\tau) \rangle_S,
\end{equation}
where
\begin{eqnarray}
\displaystyle
K(\tau) && = \sum_{i,j} (J^{-1})_{ij} \psi_i^*(\tau) \psi_j(\tau)
\nonumber \\
&& \mbox{} - \sum_i \left\{ \left \langle T_{\tau} \exp \left[ \beta
\int_{\tau}^{L} d  \tau ' \hat{f}_i (\tau ') \right] \hat{f}_i (\tau)
\right. \right. \nonumber \\
&& \hspace{0.7in} \times \left. T_{\tau} \exp \left[ \beta
\int_0^{\tau} d \tau ' \hat{f}_i (\tau ') \right] \right \rangle_0
\nonumber \\
&& \hspace{0.6in} \times \left. \left \langle T_{\tau} \exp \left[
\beta \int_0^{L} d \tau ' \hat{f}_i (\tau ') \right] \right
\rangle_0^{-1} \right\},
\end{eqnarray}
\begin{equation}
\hat{f}_i (\tau ') = \psi_i(\tau ') \hat{a}_i^{\dag}(\tau ') +
\psi_i^*(\tau ') \hat{a}_i(\tau '),
\end{equation}
\begin{equation}
\langle \bullet \rangle_S = \frac{1}{\cal Z} \int \prod_i {\cal D}
\psi_i(\tau) {\cal D} \psi_i^*(\tau) \bullet \exp [-S(\psi)],
\end{equation}
and
\begin{eqnarray}
\displaystyle
S(\psi) & = & \beta \sum_{i,j} \int_0^{L} d \tau (J^{-1})_{ij}
\psi_i^*(\tau) \psi_j(\tau) \nonumber \\
\displaystyle
& & \mbox{} - \sum_i \ln \left\langle T_{\tau} \exp \left[ \beta
\int_0^{L} d \tau \{ \psi_i(\tau) \hat{a}_i^{\dag}(\tau)
\right. \right. \nonumber \\
\displaystyle
& & \hspace{1.2in} \left. \mbox{} + \psi_i^*(\tau) \hat{a}_i(\tau) \}
\Bigg] \right\rangle_0.
\label{eq:Sfree}
\end{eqnarray}
Although Eq.\ (\ref{eq:Sfree}) may appear to be identical to Eq.\
(\ref{eq:S}), the meaning of $\langle \bullet \rangle_0$ as well as
the boundary conditions on $\psi$ have changed.  For ``boson''
boundary conditions, $\psi$ was restricted to be periodic,
i.e., $\psi_i (\tau = 0) = \psi_i (\tau = \beta L)$.  Following
Ref.~\cite{N88}, we have inserted boundary conditions appropriate to
vortices in a superconducting slab.

In the Mott insulator phase, it is again sufficient to expand the
action to quadratic order in $\psi$, yielding
\begin{equation}
S(\psi) \approx S_0(\psi) + \delta S(\psi),
\end{equation}
where $S_0(\psi)$ is given by Eq.\ (\ref{eq:quad}) and $\delta
S(\psi)$ is an extra piece due to the changed boundary conditions in a
Mott insulator with $n$ particles per site
\begin{eqnarray}
\delta S(\psi) & = & - n a_0^2 \int \frac{d^2 {\bf k} d \omega d
\omega '}{(2\pi)^4} \psi ({\bf k}, \omega) \psi^* ({\bf k}, \omega ')
\nonumber \\
&& \times \left[ \frac{e^{i \omega ' \beta L}}{(E_h + i \omega k_B T)
(E_h + i \omega ' k_B T)} \right. \nonumber \\
&& \left. \mbox{} + \frac{e^{- i \omega \beta L}}{(E_p - i \omega k_B
T) (E_p - i \omega ' k_B T)} \right].
\label{eq:brenorm}
\end{eqnarray}
We can now evaluate $\langle \hat{\cal H}_1 \rangle_{\tau}$
diagrammatically.  Denote each $\delta S(\psi)$ with a cross.  The
diagrams that give the dominant contribution to $\langle \hat{\cal
H}_1 \rangle_{\tau} - \langle \hat{\cal H}_1 \rangle_{\tau = \infty}$
as $\tau \rightarrow \infty$ are those in Fig.\ 12, i.e., those with
two external lines and an even (but nonzero) number of crosses.  In
the limit $\tau \rightarrow \infty$ we find
\begin{equation}
\langle \hat{\cal H}_1 \rangle_{\tau} - \langle \hat{\cal H}_1
\rangle_{\tau = \infty} \sim e^{- (F_p + F_h) \tau / k_B T},
\end{equation}
which is Eq.\ (\ref{eq:expdecay}).

\begin{minipage}[t]{3.2in}
\vspace{0.1in}
\epsfxsize=3.2in
\epsfbox{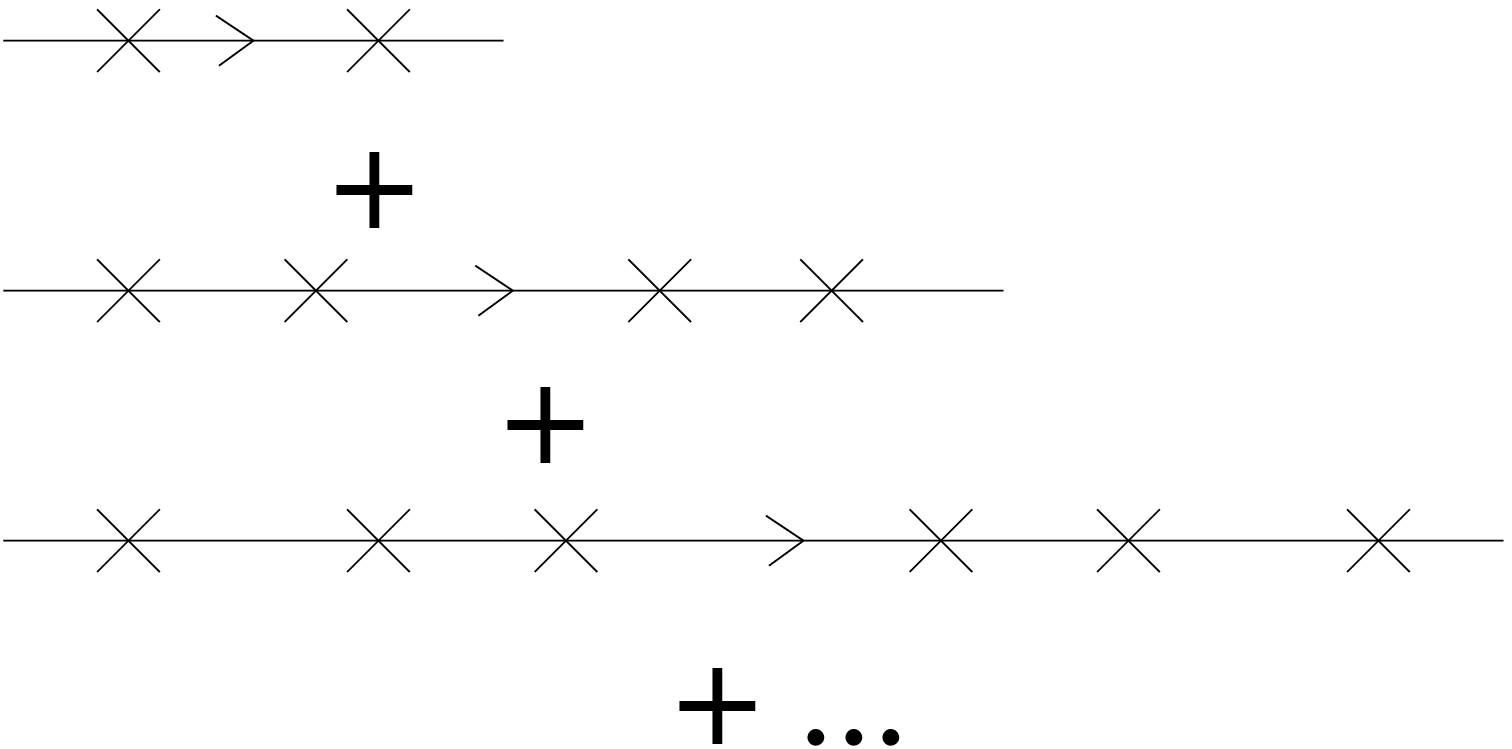}
\begin{small}
FIG.\ 12.
The diagrams that give the dominant contribution to $\langle \hat{\cal
H}_1 \rangle_{\tau} - \langle \hat{\cal H}_1 \rangle_{\infty}$ as
$\tau \rightarrow \infty$.  The crosses are free boundary insertions
as given in Eq.\ (\ref{eq:brenorm}).
\end{small}
\vspace{0.2in}
\end{minipage}

\section{Calculation of magnetic field in superfluid phase}
\label{sec-magnet}

To calculate the magnetic field in the superfluid phase, we use the
thermodynamic relation
\begin{equation}
{\bf B} = -4 \pi \frac{\partial f}{\partial {\bf H}},
\end{equation}
where $f$ is the free energy density $f = - (k_B T / Na_0^2 L_z) \ln
{\cal Z}$ and the partition function ${\cal Z}$ is given in Eq.\
(\ref{eq:Z2}) and Eq.\ (\ref{eq:S}).

We use Eq.\ (\ref{eq:Z2}) to evaluate ${\bf B}$.  The piece that
comes from ${\cal Z}_0$ is easily seen to be exactly the magnetic
field that we expect for a Mott insulator with occupation number $n$.
The overall field is
\begin{equation}
{\bf B} = \frac{n \phi_0}{a_0^2} \hat{\bf z} - \frac{4 \pi k_B T}{N
a_0^2 L_z} \left \langle \frac{\partial S}{\partial {\bf H}} \right
\rangle_S.
\end{equation}
We first define
\begin{equation}
\delta {\bf B} \equiv {\bf B} - \frac{n \phi_0}{a_0^2} \hat{\bf z}
\end{equation}
and
\begin{equation}
\delta {\bf H} \equiv {\bf H} - \frac{4 \pi \epsilon_1}{\phi_0}
\hat{\bf z}.
\end{equation}
Then Eqs.\ (\ref{eq:mu}) and (\ref{eq:h}) give
\begin{equation}
\delta {\bf H} = \frac{4\pi k_B T}{\phi_0} \left( \frac{\bf h}{a_0},
\beta \mu \right),
\end{equation}
so that
\begin{equation}
\delta {\bf B} = - \frac{\phi_0}{N a_0^2 L_z} \left( a_0 \left \langle
\frac{\partial S}{\partial {\bf h}} \right \rangle_S, k_B T \left
\langle \frac{\partial S}{\partial \mu} \right \rangle_S \right).
\label{eq:deltab}
\end{equation}

Near the transition, we can use the action of Eq.\
(\ref{eq:expandquad}) to evaluate the averages in Eq.\
(\ref{eq:deltab}), obtaining
\begin{eqnarray}
\delta {\bf B} & \approx & \frac{\beta \phi_0}{N L_z} \int \frac{d^2
{\bf k} d \omega}{(2\pi)^3} \left \langle \left| \psi ({\bf k},
\omega) \right|^2 \right \rangle \Big[ c_{\mu} \hat{\bf z} + {\bf c_h}
\nonumber \\
&& \hspace{0.5in} \mbox{} + 2 i c' \omega \hat{\bf z} - i D_{ij} (k_i
\hat{e_j} + k_j \hat{e_i}) \Big] \nonumber \\
&& \mbox{} - \frac{\phi_0 k_B T}{N a_0^2 L_z} \frac{\partial
u}{\partial \mu} \hat{\bf z} \int d^2 {\bf r} d \tau \left \langle
\left| \psi ({\bf r}, \tau) \right|^4 \right \rangle + \cdots, 
\label{eq:quadexp}
\end{eqnarray}
where the omitted terms reflect contributions left out of Eq.\
(\ref{eq:expandquad}).  [Note that $\partial u / \partial \mu$ is
proportional to the coefficient of the $\omega |\psi|^4$ term that
would appear in Eq.\ (\ref{eq:expandquad}).]  To see which terms will
yield the most important contributions, we transform the axes of space
and imaginary time as in Appendix~\ref{sec-rg}.  We have already
determined the scaling of the coefficients under the action of the
renormalization group, because they are the same coefficients as
appear in the action.  We discard those terms whose coefficients are
irrelevant, and obtain
\begin{equation}
\delta {\bf B} = \frac{\beta \phi_0}{N L_z} \int \frac{d^2 \tilde{\bf
k} d \tilde{\omega}}{(2\pi)^3} \left \langle \left| \psi (\tilde{\bf
k}, \tilde{\omega}) \right|^2 \right \rangle \left[ {\bf v} - 2 i D
\tilde{\bf k} \right],
\label{eq:mnresdb}
\end{equation}
where ${\bf v}$ is given by Eq.\ (\ref{eq:angle}).  We then iterate
the renormalization group until $|r(l)| = 1$, so that we are far from
the transition, and can use mean field theory.  If the flow begins on
the Mott insulator side of the transition, then we will iterate until
we are deep in the Mott insulator phase and Eq.\ (\ref{eq:mnresdb})
vanishes.  If the flow begins on the superfluid side of the
transition, then deep in the superfluid phase we can set $|\psi
(\tilde{\bf r}, \tilde{\tau})| \approx \sqrt{\beta a_0^2 r(l)/2u(l)}$,
independent of space and imaginary time.  This yields
\begin{equation}
\delta {\bf B} \approx \phi_0 \beta a_0^2 {\bf v} \langle |\psi
|^2 \rangle.
\label{eq:rawdb}
\end{equation}
We therefore identify $\hat{\bf v}$ with the average direction of the
tilted defects, and $n = \beta a_0^2 |{\bf v}| \langle |\psi |^2
\rangle$ with their density.  Performing a scaling analysis similar to
what was done for the condensate fraction $n_0$ in
Appendix~\ref{sec-rg}, we obtain (away from the multicritical point)
\begin{equation}
\delta {\bf B} \approx \phi_0 \hat{\bf v} \frac{|r \ln r|}{16\pi D}
\end{equation}
using $C = |{\bf v}|$ and with $D$ given in Eq.\ (\ref{eq:D}).  At
the multicritical point, ${\bf v} = 0$, so Eq.\ (\ref{eq:rawdb})
gives $\delta {\bf B} = 0$ near the transition.

\end{multicols}

\begin{references}
\bibitem{Note}  We concentrate here on a simplified ``tight binding''
model which assumes the energy cost of double occupancy is less than
that required for one vortex to wander in the interstitial region.
The energy cost of multiple occupancy could be lowered into this
regime by increasing cross sectional area of the defects.
\end{references}
\end{document}